\documentclass[lettersize,journal]{IEEEtran}
\usepackage{amsmath,amsfonts}
\usepackage{algorithmic}
\usepackage{algorithm}
\usepackage{array}
\usepackage{textcomp}
\usepackage{stfloats}
\usepackage{url}
\usepackage{verbatim}
\usepackage{graphicx}
\usepackage{cite}
\usepackage{hyperref}
\usepackage{float}
\usepackage{cleveref} 
\usepackage{subfigure}
\usepackage{balance}
\usepackage{bm}
\usepackage{multirow}
\usepackage{booktabs}
\usepackage{color}   
\usepackage{xcolor}
\usepackage{gensymb}
\usepackage[commandnameprefix=always,final]{changes}
\usepackage[utf8]{inputenc}
\makeatletter
\newcommand*{\new}{\@ifnextchar\bgroup{\new@}{\textcolor{black}}}
\newcommand*{\new@}[1]{{\textcolor{black}{#1}}}
\makeatother
\renewcommand{\arraystretch}{1.25} 

\title{T-ADD: Enhancing DOA Estimation Robustness Against Adversarial Attacks} 

\author{Shilian~Zheng,
        Xiaoxiang~Wu,
        Luxin~Zhang,
        Keqiang~Yue, 
        Peihan~Qi and
        Zhijin~Zhao
\thanks{\textcolor{black}{This work was supported in part by the National Basic Scientific Research
of China under Grant JCKY2023110C099 and the Key R\&D Program of China under Grant 2025YFG0100400. \textit{(Corresponding author: Shilian Zheng.)}}}
\thanks{S. Zheng is with the College of Communication Engineering, Hangzhou Dianzi University, Hangzhou 310018, China. He is also with No. 11 Research Center, National Key Laboratory of Electromagnetic Space Security, Jiaxing 314033, China (e-mail: lianshizheng@126.com).}
\thanks{X. Wu, and Z. Zhao are with the College of Communication Engineering, Hangzhou Dianzi University, Hangzhou 310018, China (e-mail: douckze@outlook.com, zhaozj03@hdu.edu.cn).

K. Yue is with Key Laboratory of RF Circuits and Systems, Ministry of Education, Hangzhou Dianzi University, Hangzhou 310018, China (e-mail: kqyue@hdu.edu.cn).

L. Zhang is with  No. 11 Research Center, National Key Laboratory of Electromagnetic Space Security, Jiaxing 314033, China (e-mail: lxzhangMr@126.com).

P. Qi is with the State Key Laboratory of Integrated Service Networks, Xidian University, Xi’an 710071, China (e-mail:  phqi@xidian.edu.cn).
}
}

\begin{document}
\maketitle
\begin{abstract}

Deep learning has achieved remarkable success in direction-of-arrival (DOA) estimation. However, recent studies have shown that adversarial perturbations can severely compromise the performance of such models. To address this vulnerability, we propose Transformer-based Adversarial Defense for DOA estimation (T-ADD), a transformer-based defense method designed to counter adversarial attacks.
To achieve a balance between robustness and estimation accuracy, we formulate the adversarial defense as a joint reconstruction task and introduce a tailored joint loss function. Experimental results demonstrate that, compared with three state-of-the-art adversarial defense methods, the proposed T-ADD significantly mitigates the adverse effects of widely used adversarial attacks, leading to notable improvements in the adversarial robustness of the DOA model.
\end{abstract}


 \section{Introduction}

\IEEEPARstart{D}{irection-of-Arrival} (DOA) estimation is widely applied in fields such as multiple-input multiple-output (MIMO) radar systems \cite{2010mimodoa} and autonomous driving \cite{wan2021autodrive}, to provide critical spatial information for complex environmental perception and decision-making. Traditional DOA estimation methods, including subspace-based methods \cite{schmidt1986multiple,roy1989esprit} and beam-forming \cite{krim1996two}, perform well under ideal conditions, but these approaches are highly sensitive to the number of snapshots and the signal-to-noise ratio (SNR), often resulting in performance degradation in practical scenarios. 
The maximum likelihood (ML) method \cite{1990ml} provides theoretically optimal estimation accuracy, but its high computational complexity significantly limits its applicability in real-time systems.

The advancement of MIMO technology has significantly improved the angular resolution of \textcolor{black}{1D} DOA \cite{huang2018doamimoDL} \textcolor{black}{and 2D DOA \cite{20252ddoamajor} estimation}, but it also raises concerns regarding the high cost of RF chains and computational resources. To address this issue, Zhang et al. \cite{zhang2022refdoa} proposed a dynamic ML algorithm applicable to various array structures, which effectively reduces computational complexity. Moreover, the multidimensional nature of MIMO systems has motivated the use of tensor-based modeling techniques. For instance, Kwan et al. \cite{zhang2022refmimochanneltarget} employed tensor decomposition to achieve joint estimation of DOA and Doppler parameters. 

Deep learning, as a powerful machine learning technique, extracts signal features from complex environments and offers higher accuracy for DOA estimation compared to traditional methods\cite{papageorgiou2021deep,zheng2024doa}.
Some researchers have attempted to integrate classic subspace-based methods with deep learning to leverage both interpretability and performance advantages \cite{merkofer2023damusic, wu2024gridless}. Wu et al. \cite{wu2024gridless} employed neural networks to reconstruct the covariance matrix, thereby reducing the training sample requirements for multi-source DOA estimation. \textcolor{black}{Ji et al. \cite{transmusic} proposed a Transformer-enhanced MUSIC framework for superior DOA estimation under low-resolution ADC quantization. }
Studies have further simplified the traditional estimation pipeline by directly using the covariance matrix as input for end-to-end DOA estimation \cite{papageorgiou2021deep, dctvit}. 
Papageorgiou et al. \cite{papageorgiou2021deep} employed the covariance matrix as the input to a CNN.
Guo et al. \cite{dctvit} used the covariance matrix as input and integrated it with a Vision Transformer to achieve robust performance for multiple signal sources under low SNR conditions.
However, the computation of covariance matrices inevitably leads to information loss. To address this, some studies have focused on directly utilizing the raw in-phase and quadrature (IQ) data to fully exploit the available signal information. Chen et al. \cite{sdoa} improved DOA estimation under imperfect array conditions by using IQ data input. Zheng et al.~\cite{zheng2024doa} proposed neural network-based classification and regression approaches to estimate discrete and continuous angles, respectively, which maintain stable performance even in extreme multi-source DOA estimation scenarios. \textcolor{black}{Wang et al. \cite{wang2025trans2}  proposed a Transformer-based DOA estimation method with the antenna-oriented attention mechanism, aiming to enhance performance in multi-source and array imperfection scenario. 
Although Transformer architectures have been utilized in DOA estimation \cite{transmusic, wang2025trans2}, these approaches did not specifically address adversarial threats and thus differ fundamentally from our research focus on adversarial defense for DOA estimation.}

Despite the superior performance of deep learning methods in DOA estimation, their vulnerability to adversarial attacks has raised significant concerns \cite{madry2019deeplearningmodelsresistant}. Adversarial attacks refer to the intentional injection of imperceptible perturbations into input data, which can lead neural networks to make incorrect predictions. Depending on the attacker's level of access to the target model, adversarial attacks are typically categorized into white-box and black-box scenarios.
In white-box scenarios, the attacker is assumed to have full access to the model, including its architecture and parameters. White-box attack methods include gradient-based approaches such as Projected Gradient Descent (PGD) \cite{madry2019deeplearningmodelsresistant} and Momentum Iterative Method (MIM) \cite{dong2018boosting}, as well as optimization-based strategies like DeepFool \cite{moosavi2016deepfool}.
In contrast, black-box attacks are conducted without knowledge of the model’s internal structure or parameters, relying solely on the model’s predictions. These attacks are generally divided into two categories: query-based attacks \cite{chen2020hopskipjumpattack}, which iteratively query the model to craft adversarial samples, and transfer-based attacks \cite{wu2024transferattack}, which generate perturbations under the assumption of adversarial transferability.

In the fields of communications and signal processing, adversarial defense techniques can generally be categorized into three main types: input-level defenses, model-level defenses, and detection-based defenses \cite{wang2023advatkdefsurvey}.

Input-level defenses aim to enhance model robustness by applying preprocessing techniques that suppress adversarial perturbations within the input. For example, Lou et al. \cite{zheng2021puaa} proposed an LSTM-based data reconstruction method to effectively counteract adversarial perturbations in spectrum sensing. Zhou et al. \cite{zhou2024ganbasedsiamese} introduced a siamese Generative Adversarial Networks (GAN) framework to improve the robustness of automatic modulation classification (AMC) models. Lin et al. \cite{lin2024dibad} developed a defense strategy based on the information bottleneck principle to isolate adversarial components, thereby enhancing the model’s resistance. Qi et al. \cite{qi2024wavedesigndef} proposed an embedded waveform design defense using GANs to protect wireless communication links from adversarial attacks. Yu et al. \cite{zhang2024hfad} applied homomorphic filtering to suppress high-frequency components in adversarial samples, improving AMC robustness against both white-box and black-box attacks.
Model-level defenses typically include adversarial training and defensive distillation techniques. Sahaya et al. \cite{sahay2022ensembleamcdef} employed ensemble adversarial training across multiple models, significantly boosting the robustness of AMC systems under various attack scenarios. Roli et al. \cite{Roli2023transdistillation} utilized a transformer-based teacher network with high robustness, combined with attention map-guided knowledge distillation to enhance the defense performance. Chen et al. \cite{chen2024learntodefend} further introduced a dual-teacher framework, where one teacher focuses on robustness and the other maintains clean sample performance, thereby improving the generalization capability of AMC models under varying input distributions.
Detection-based defenses focus on identifying and filtering potential adversarial samples, often by measuring discrepancies between features before and after perturbation. In the context of Internet of Things (IoT) intrusion detection, Jiang et al. \cite{JIANG2022194} proposed a detection framework that integrates strategic feature grouping and multi-model fusion, significantly improving robustness against adversarial intrusion behaviors.

In recent years, transformer have gained significant attention due to their powerful self-attention mechanisms for global feature modeling, and have been widely applied to tasks such as image classification and reconstruction \cite{liu2022swin, li2023efficientgrl}. In the field of adversarial defense, the potential of transformers is also being increasingly explored. For instance, Mounim et al. \cite{mounim2023transjingmai} employed a transformer-based GAN architecture—utilizing transformers for both the generator and discriminator—to reconstruct adversarially perturbed finger-vein images and suppress adversarial noise. Similarly, Yan et al. \cite{yan2025transgan} enhanced the robustness of indoor localization systems using a transformer-based GAN framework. 

Although adversarial defense techniques have made notable progress in wireless communication areas, research efforts targeting DOA estimation remain limited. Yang et al. \cite{yangzhuangattack} demonstrated that adversarial perturbations could deceive DOA estimation models, resulting in estimation errors. Thus, measures are needed to defend against adversarial attacks. However, comprehensive adversarial defense studies tailored to DOA estimation are still scarce. To best of our knowledge, the notable attempt is by Xu et al. \cite{xu2024evaporativegan}, who proposed a GAN-based preprocessing module to mitigate the effects of coherent jamming attacks on DOA estimation. Nonetheless, their work primarily addresses coherent source interference scenarios and does not consider a broader range of adversarial attack targeting deep learning models. In practical systems, adversarial perturbations are often unknown and may vary over time. This highlights the pressing need for generalizable defense strategies that can enhance DOA robustness under diverse attack conditions.

To tackle the aforementioned challenges, we introduce the Transformer-based Adversarial Defense for DOA estimation (T-ADD), designed to learn the complex mapping between adversarial and clean samples,eliminating perturbations in the input data thereby enhancing the model’s generalization and robustness performance under varying attack types. 
Our main contributions are as follows:

\begin{itemize}
    \item \textcolor{black}{While prior works have applied GAN to counter coherent interference or used transformer for end-to-end and model-based DOA estimation, this is the first work to leverage a transformer-based network specifically for adversarial defense in DOA estimation across various attack types.} A novel defense framework, termed T-ADD, is proposed to enhance the robustness of DOA estimation under adversarial attacks.  
    \item A joint reconstruction loss function is designed to simultaneously preserve the structural integrity of the original signal and suppress adversarial perturbations. Ablation studies are conducted to evaluate the impact of different hyperparameters. 
    \item Extensive experiments are conducted to compare the proposed method with three state-of-the-art adversarial defense techniques under various representative attack scenarios. Results demonstrate that the proposed approach effectively improves adversarial robustness while maintaining high estimation accuracy on clean inputs.
    \end{itemize}
    
\section{{\textcolor{black}{Preliminaries}}}
\subsection{{Signal Model}}
We consider of two array types including uniform linear array (ULA) and sparse linear array (SLA). Firstly, When examining a uniformly spaced linear array consisting of \(M\) elements, each separated by a distance $\Delta$, the independent far-field narrow-band signal \( s_1,s_2,...,s_l \) received from different directions \( \theta_1, \theta_2, \dots, \theta_L \). The signal \textcolor{black}{received} by the \( m \)-th element of the array is given by the following expression:

\begin{equation}
    x_m(n) = \sum_{l=1}^L s_l(n)e^{-j2\pi \Delta (m-1)\sin\theta_l / \lambda} + \mathcal{{N}}_m(n),
\end{equation}
where \( \lambda \) is the wavelength, and $ \mathcal{N}_m(n) $ represents independent Gaussian noise. The vector form of the received signal is:
\begin{equation}
    \mathbf{x}(n) = \mathbf{A}(\boldsymbol{\theta})\mathbf{s}(n) + \mathbf{N}(n), 
\end{equation}
where \( \mathbf{s}(n) = [s_1(n), \dots, s_L(n)]^T \) is the signal vector,  \( \mathbf{N}(n) = [\mathcal{{N}}_1(n), \dots, \mathcal{{N}}_M(n)]^T \) is the noise vector, and \( \mathbf{A}(\boldsymbol{\theta}) = [\mathbf{a}(\theta_1), \dots, \mathbf{a}(\theta_L)] \) is the steering matrix. Here \( \mathbf{a}(\theta_l) \) is the steering vector of the \( l \)-th signal, given by:

\begin{equation}
    \mathbf{a}(\theta_l) = [1, e^{-\phi(\theta_l)}, \dots, e^{-(M-1)\phi(\theta_l)}]^T,
\end{equation}
where \( \phi(\theta_l) = j2\pi d \sin\theta_l / \lambda \). 

When the array samples \( K \) snapshots, the signal can be represented in matrix form as:
\begin{equation}
    \mathbf{X} = \mathbf{A}\mathbf{S} + \boldsymbol{N}, 
\end{equation}
where \( \mathbf{X} \) is the received signal matrix, \( \mathbf{S} \) is the signal matrix, and \( \boldsymbol{N} \) is the noise matrix. The sample covariance matrix $\mathbf{\hat{R}}_{xx}$ is then computed via
\begin{equation}
    {{\hat{\bf R}}_{xx}} = \frac{1}{K}\sum\limits_{n = 1}^K {\mathbf{x}} (n){{\mathbf{x}}^{\rm{H}}}(n).
\end{equation}

For SLA, the array elements are typically distributed in a non-uniform manner, i.e., the inter-element spacing is no longer uniform. Let the array position vector be defined as $\boldsymbol{\Delta} = [\Delta_1,\Delta_2,\dots,\Delta_M]^T$, where $\Delta_m$ denotes the normalized position (in units of half-wavelength $\lambda/2$) of the $m$-th element along the array axis. The array steering matrix requires correction and is reformulated as:
\begin{equation}
\mathbf{A}_{\text{NU}}(\boldsymbol{\theta}) = 
\begin{bmatrix}
    e^{-j\pi \Delta_1 \sin\theta_1} & \cdots & e^{-j\pi \Delta_1 \sin\theta_L} \\
    \vdots & \ddots & \vdots \\
    e^{-j\pi \Delta_M \sin\theta_1} & \cdots & e^{-j\pi \Delta_M \sin\theta_L}.
\end{bmatrix}
\end{equation}
The covariance matrix of the received signal can then be modeled as:
\begin{equation}
    \mathbf{R}_{xx}^{\text{NU}} = \mathbb{E}[\mathbf{x}(n)\mathbf{x}^\text{H}(n)] = \mathbf{A}_{\text{NU}} \mathbf{R}_{ss} \mathbf{A}_{\text{NU}}^\text{H} + \sigma^2 \mathbf{I}_M,
\end{equation}
where $\mathbf{R}_{ss} = \mathbb{E}[\mathbf{s}(n)\mathbf{s}^\text{H}(n)]$ is the source signal covariance matrix, $\sigma^2$ is the noise power, and $\mathbf{I}_M$ denotes the $M \times M$ identity matrix. Similar to uniform linear array, for SLA, the sample covariance matrix can be estimated via sample averaging after collecting $K$ snapshots. The sample covariance matrix is defined as:
\begin{equation}
    \hat{\mathbf{R}}_{xx}^{\text{NU}} = \frac{1}{K}\sum_{n=1}^{K} \mathbf{x}(n)\mathbf{x}^\text{H}(n).
\end{equation}
Since the sample covariance matrix is the widely-used input format, the DOA estimation model employed for adversarial defense evaluation in this study adopts the network architecture proposed by~\cite{papageorgiou2021deep}.
\subsection{Threat Model}
\subsubsection{Definition of Adversarial Attack}
Adversarial attacks aim to craft inputs that mislead a target model into making incorrect predictions, while keeping the added perturbations imperceptible or bounded within a specified $\ell_p$-norm ball.
Given an input $\bm{x}$ and its corresponding ground-truth label $y$, let $f(\cdot)$ denote the prediction function of a deep learning model, such as the CNN proposed by\cite{papageorgiou2021deep}. An adversarial sample is defined as:
\begin{equation}
    \bm{x}_{{adv}} = \bm{x} + \boldsymbol{\delta}, \quad \text{subject to } \|\boldsymbol{\delta}\|_{p} \leq \varepsilon,
\end{equation}
where $\boldsymbol{\delta}$ is the perturbation constrained by a norm bound $\varepsilon$.

To formally assess attack success, an adversarial indicator function $\mathbb{I}$ is defined as:
\begin{equation} \label{eq:adv-id}
    \mathbb{I}(f(\bm{x}_{{adv}}) \neq y) = 
    \begin{cases}
        1, & \text{if } f(\bm{x}_{{adv}}) \neq y \\
        0, & \text{otherwise}
    \end{cases}
\end{equation}

In gradient-based attacks, the perturbation $\boldsymbol{\delta}$ is optimized to maximize the loss function $\mathcal{L}(f(\bm{x} + \boldsymbol{\delta}), y)$ under the given constraint, thereby increasing the likelihood that the adversarial input leads to a misclassification.

To ensure the imperceptibility of adversarial samples, it is essential to effectively control the energy of the perturbation. The Signal-to-Interference Ratio (SIR) is defined as:
\begin{equation}
    \mathrm{SIR} = \frac{P_{\mathrm{clean}}}{P_{\mathrm{pert}}},
\end{equation}
where $P_{\mathrm{clean}}$ and $P_{\mathrm{pert}}$ denote the power of the clean signal and the perturbation $\bm{\delta}$. 
To regulate the perturbation energy, the adversarial perturbation can be scaled accordingly as follows:
\begin{equation}
     \bm{x}_{adv}' =  \bm{x}_{adv} \times \sqrt{ \frac{P_{\mathrm{clean}}}{P_{\mathrm{pert}}} \times 10^{\mathrm{SIR} / 10} }.
\end{equation}
\subsubsection{Attack Method}
Gradient-based white-box adversarial attacks can be broadly categorized into one-step and multi-step iterative methods (e.g., PGD, MIM). Among these, multi-step iterative approaches are widely adopted due to their high attack success rates and robustness.

The PGD attack generates adversarial samples by performing multiple small-step updates, each followed by a projection back onto the allowed perturbation region. The update rule at each iteration is given by:
\begin{equation}
    \bm{x}_{i+1} = \Pi_{\bm{x} + \varepsilon} 
    \left( 
        \bm{x}_i + \psi \cdot \operatorname{sign}\left( \nabla_{\bm{x}} \mathcal{L}(f(\bm{x}_i), y) \right) 
    \right),
\end{equation}
where $\Pi_{\bm{x}+\varepsilon}(\cdot)$ denotes the projection operator onto the $\ell_{\infty}$-bounded neighborhood centered at $\bm{x}$, $\psi$ is the step size, and $\mathrm{sign}(\cdot)$ is the sign function.

The MIM enhances the iterative update by introducing a momentum term, which accumulates past gradients to mitigate oscillations and escape local optima, thereby improving attack efficiency. The update rule is given by:
\begin{equation}
    \bm{x}_{i+1} = \bm{x}_i + \psi \cdot \text{sign}(g_{t+1}),
\end{equation}
where $g_{i+1}$ represents the accumulated gradient at iteration $i+1$, updated as:
\begin{equation}
    g_{i+1} = \omega \cdot g_i + \frac{\nabla_{\bm{x}} \mathcal{L}(f(\bm{x}_i), y)}{\|\nabla_{\bm{x}} \mathcal{L}(f(\bm{x}_i), y)\|_1},
\end{equation}
where $\omega$ is the momentum decay factor that smooths the gradient updates, and $\|\cdot\|_1$ denotes the $\ell_1$-norm.

\subsection{Defense Model}
The goal of adversarial defense in this study is to enhance the robustness of DOA estimation models by reducing the impact of adversarial perturbations. 
\textcolor{black}{Adversarial robustness is defined as the model’s resistance to adversarial attacks~\cite{carlini2019evaluatingroubustness}, and can be formally expressed as: : 
\begin{equation}
    \mathbb{E}_{(\bm{x}, y) \sim \mathcal{X}}\left[\max_{\bm{\delta} \in \Omega} \mathcal{L}\left(f\left(\bm{x}+\bm{\delta}\right), y\right)\right],
\end{equation}
}where \textcolor{black}{$\mathbb{E}(\cdot)$ denotes the mathematical expectation,} $\Omega$ denotes the space of admissible adversarial perturbations.

Instead of modifying the model architecture or employing adversarial training, we adopt a strategy that preprocesses input data to attenuate \(\bm{\delta}\). To \textcolor{black}{achieve this}, we introduce a signal reconstruction module $T(\cdot)$ that aims to recover the clean representations of adversarial samples prior to inference. The defense objective can be formally expressed as:\textcolor{black}{
\begin{equation}\label{eq:def-goal}
    \min_{T} 
    \mathbb{E}_{(\bm{x}, y) \sim \mathcal{X}}
    \left[
        \max_{\bm{\delta} \in \Omega}
        \mathcal{L}\big(f(T(\bm{x}+\bm{\delta})), y\big)
    \right].
\end{equation}}This formulation ensures that, after reconstruction, the adversarial samples \textcolor{black}{incur minimal distortion relative to the original clean samples}, thereby preserving the model's \textcolor{black}{predictive} integrity against adversarial attacks.


\section{Defense Method}
\label{sec:defense}
Direct optimization of Eq.~\eqref{eq:def-goal} over all possible perturbations $\bm{\delta} \in \Omega$ is generally intractable due to the high dimensionality and non-linearity of the adversarial space. 
To address this issue, we assume that the clean sample $\bm{x}_n$ approximately minimizes the empirical risk on the original model. Under this assumption, the reconstruction function is expected to satisfy the following two constraints: (i) when the input is clean, the reconstruction should preserve the original structure, i.e., $T(\bm{x}_n) \approx \bm{x}_n$; (ii) when the input is adversarial, the reconstruction should suppress the perturbations and approximate the corresponding sample, i.e., $T(\bm{x}_{{adv}}) \approx \bm{x}_n$. These constraints enable the defense mechanism to operate without prior knowledge of the specific attack and generalizes to unseen perturbations at inference time.

\subsection{T-ADD Framework Overview}\label{subsec:framework}
\begin{figure*}[tbp]
  \centering
  \includegraphics[width=0.90125\linewidth]{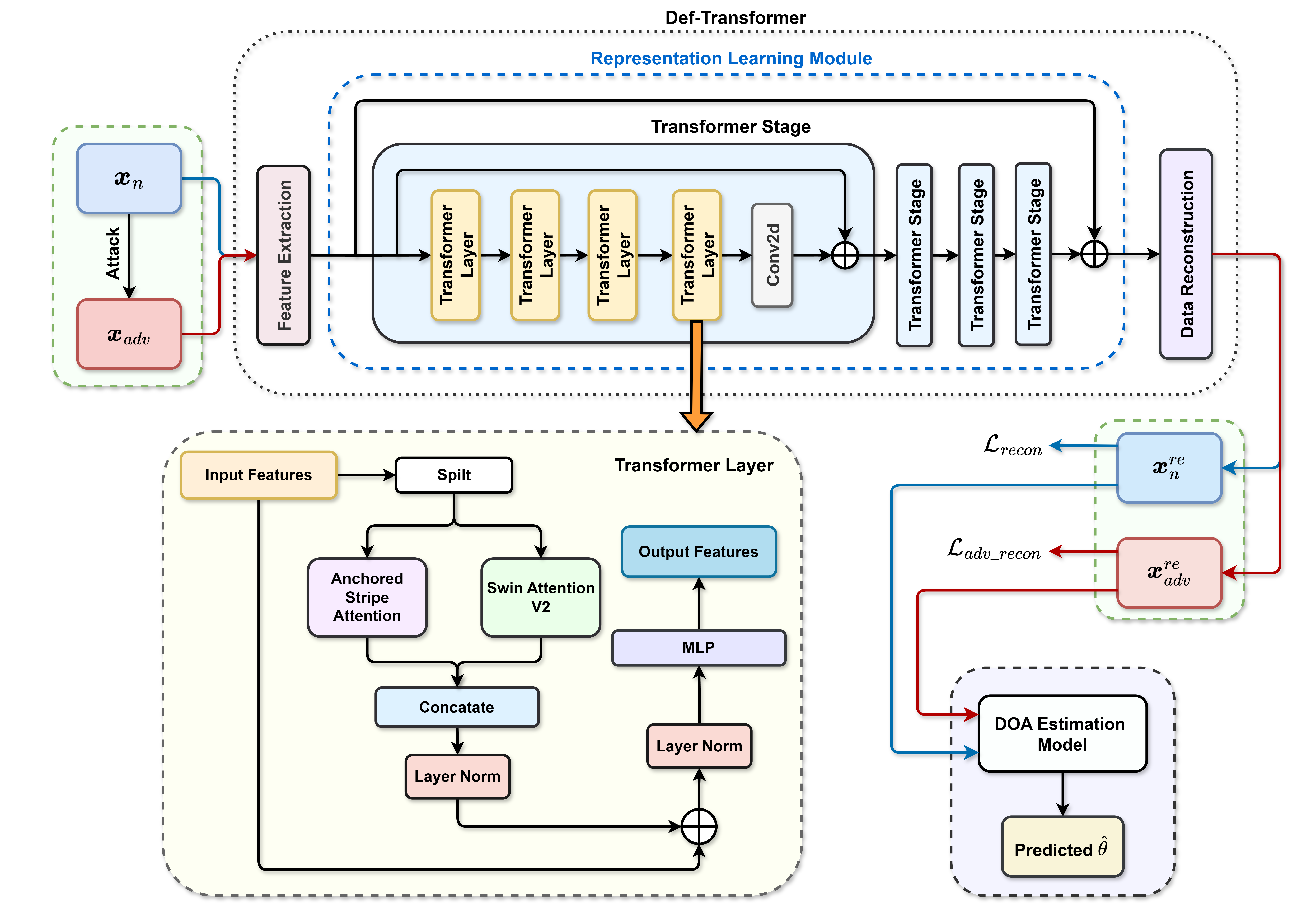}
  \caption{T-ADD Overall Framework. During the training phase of the Def-Transformer, empirical risk minimization (ERM) is performed using a joint sample reconstruction loss composed of the reconstruction loss $\mathcal{L}_{{recon}}$ and the adversarial reconstruction loss $\mathcal{L}_{{adv\_recon}}$. The resulting reconstructed samples effectively suppress vulnerable adversarial perturbations, thereby enhancing the robustness of DOA estimation. In the inference phase, the DOA estimation model operates on the reconstructed samples for angle prediction.}
  \label{fig:raden}
\end{figure*}

To achieve the above objectives, we propose a transformer-based adversarial defense framework named T-ADD. As illustrated in \textcolor{black}{Fig.~\ref{fig:raden}}, the overall pipeline consists of two main components: the Def-Transformer and a baseline DOA estimation model. The reconstruction can be described using the following equation:

\begin{equation}
    \tilde{\bm{x}}^{\textcolor{black}{re}} = \mathcal{T}_{\xi}(\tilde{\bm{x}}),
    \label{eq:reconstruction}
\end{equation}
where $\tilde{\bm{x}} \in \{ \bm{x}_n, \bm{x}_{adv} \}$, and $\mathcal{T}_{\xi}(\cdot)$ denotes the Def-Transformer network parameterized by $\xi$.

As illustrated in \textcolor{black}{Fig.~\ref{fig:raden}}, the defense framework comprises a trained Def-Transformer and a DOA estimation model. Input samples $\tilde{\bm{x}}$, either clean \textcolor{black}{\(\bm{x}_n\)} or adversarial \textcolor{black}{\(\bm{x}_{adv}\)}, are first processed by the Def-Transformer to generate reconstructed samples. These reconstructed samples are then fed into the DOA estimation model. This defense mechanism ensures robustness against adversarial attacks while maintaining low estimation error. 

During training, the Def-Transformer is trained on paired datasets of clean and adversarial samples. It produces reconstructed $\tilde{\bm{x}}^{\textcolor{black}{re}}$ through forward propagation. The original clean samples \(\bm{x}_n\) are used as targets to ensure consistent reconstruction of both clean and adversarial inputs, thereby enhancing the model's ability to handle various input scenarios effectively. It is worth noting that only one type of adversarial sample is used during training.

\subsection{Def-Transformer Design}
\subsubsection{Main Architecture}
CNNs primarily focus on local feature representations, which limits their ability to capture global dependencies and spatial positional relationships among components. In contrast, transformer architectures have demonstrated strong capabilities in modeling long-range dependencies through attention mechanisms. The Def-Transformer network design is illustrated in \textcolor{black}{Fig.~\ref{fig:raden}}. Motivated by computational complexity and reconstruction fidelity considerations, we adopt a small-scale architecture with an embedding dimension of 64. Given input data, the Def-Transformer processes it and outputs the purified samples.  The input dimension is consistent with that of the baseline DOA estimation model, formatted as $3\times M \times M$.

The entire Def-Transformer can be divided into three main parts: the feature extraction unit, the representation learning module, and the data reconstruction unit.
The feature extraction module processes the input data through convolution operations to generate initial feature maps. These feature maps include basic sample information, serving as the foundation for subsequent deep learning processing. Specifically, this module comprises a convolutional layer with a kernel size of $3 \times 3$ and 64 output channels.

The representation learning module consists of four transformer stages, designed to capture both global and local information from the input data. Each stage contains four transformer layers. This module concludes with a convolutional layer (kernel size $3 \times 3$) that integrates and outputs the final high-level feature representation. To concurrently attain the objectives of retaining both shallow and deep feature hierarchies, skip connections are incorporated, allowing shallow and deep features to be jointly utilized in the final representation. 

After passing through multiple transformer layers, the generated feature maps will be overlaid with the initial feature maps and passed to the data reconstruction unit to restore the target sample from the learned features. Specifically, this module consists of a single convolutional layer with a kernel size of $3\times3$ and 3 output channels.

\subsubsection{Transformer Layer}
Transformer layer is the key component of the transformer stage, that is illustrated in \textcolor{black}{Fig.~\ref{fig:raden}}. In each layer, both window-based self-attention and Anchored Stripe Self-Attention (ASSA) mechanisms are incorporated to model hierarchical local and global feature dependencies. The attention window size is set to $8\times8$. The transformer layer consists of parallel branches: the first branch utilizes parallel self-attention modules to enhance global modeling capability; the \textcolor{black}{second} branch performs feature fusion through skip connections.

In the ASSA, an anchor set $\mathcal{A}$ is introduced as an auxiliary representation for the query ($\mathbf{Q}$), key ($\mathbf{K}$), and value ($\mathbf{V}$) triplets to enable sparse attention modeling. These anchors are designed as a sparse representation of the input feature map, facilitating indirect similarity propagation. This approach not only preserves critical information but also significantly reduces computational overhead. The  operation principle of ASSA is as follows:

\begin{equation} \label{eq:anchor-attention}
\begin{aligned}
\mathbf{Y} &= \mathbf{M}_e \cdot (\mathbf{M}_d \cdot \mathbf{V}), \\
\mathbf{M}_e &= \operatorname{Softmax}\left(\frac{\mathbf{Q} \mathcal{{A}}^T}{\sqrt{d}}\right), \\
\mathbf{M}_d &= \operatorname{Softmax}\left(\frac{\mathcal{{A}} \mathbf{K}^T}{\sqrt{d}}\right),
\end{aligned}
\end{equation}
where, $\mathbf{Q}$, $\mathbf{K}$, and $\mathbf{V}$ denote the query, key, and value matrices, respectively, $\mathcal{A}$ represents the anchor matrix, \(\text{Softmax}(\cdot)\) is the softmax function that applied row-wise and $d$ denotes the dimensionality of each token. Since the number of anchors is smaller than the total number of tokens, the computational cost of the attention maps $\mathbf{M}_e$ and $\mathbf{M}_d$ is significantly lower than that of the full global attention map $\mathbf{M}$.

To further reduce computational complexity, and considering the redundancy in features extracted by global self-attention mechanisms, the ASSA incorporates a stripe-based attention mechanism, which restricts local attention computation to horizontal and vertical stripes, including their shifted versions. These four stripe patterns are applied alternately to reduce redundant modeling while preserving the capacity for global context modeling, thereby significantly lowering the computational burden of attention.

In the ASSA, downsampling of the input feature map is achieved during the anchor projection stage by applying average pooling followed by a linear transformation. The downsampling factor is denoted as $\mathcal{S}$ and is set to an integer no less than 2.

Based on the above design, the resulting ASSA mechanism exhibits a computational complexity of

\begin{equation}
\mathcal{O}\left(\left(3 + \frac{2}{\mathcal{S}^2}\right) \cdot \eta \cdot C^2 + \left(1 + \frac{2}{\mathcal{S}^2}\right) \cdot \eta^2 \cdot C\right),
\end{equation}
where $\eta$ denotes the product of the spatial dimensions of the input feature map, and $C$ is the number of feature channels. In contrast, the computational complexity of global self-attention in ViT~\cite{conde2022swin2sr} is given by

\begin{equation}
\mathcal{O}(4\eta \cdot C^2 + 2\eta^2 \cdot C).
\end{equation}

It can be observed that the proposed ASSA mechanism significantly reduces the computational cost, especially on large-scale feature maps, thus offering improved efficiency over conventional global attention methods.

Additionally, the transformer layer also integrates the Swin transformer V2 \cite{liu2022swin} window attention mechanism. 
This mechanism processes feature maps using a hierarchical structure and a post-normalization strategy. Additionally, it replaces the conventional dot-product attention with a scaled cosine similarity function in the attention computation, thereby enhancing the representational capability. The calculation formula is as follows:
\begin{equation}
\mathbf{M}_v = {\text{Softmax}} \left( \frac{\cos(\mathbf{Q}, \mathbf{K}^T)}{\gamma } \cdot\upsilon\right) \mathbf{V}, 
\end{equation}
where $\gamma$ is a learnable scaling factor, and $\upsilon$ denotes the relative positional bias derived from positional encodings.

\subsection{Loss Function}
\label{subsec:loss}
To enable the model to distinguish between clean samples and adversarial samples effectively, it must not only eliminate adversarial perturbation but also \textcolor{black}{maintain consistent performance } the performance on clean samples. 

\textcolor{black}{In this paper, the adversarial robustness objective is approximated through a joint reconstruction loss that balances the fidelity of clean samples and the robustness against adversarial perturbations.} The total training loss is formulated as follows:

\begin{equation}
    \mathcal{L}_{\text{Def}} = \alpha\cdot\underbrace{\|\mathcal{T}_{\xi}(\bm{x}_n) - \bm{x}_n \|_1}_{\mathcal{L}_{recon}} \;+\;\,\underbrace{\|\mathcal{T}_{\xi}(\bm{x}_{adv}) - \bm{x}_n\|_1}_{\mathcal{L}_{adv\_recon}},
    \label{eq:def_loss}
\end{equation}
where $\alpha$ is the weight parameter of the clean reconstruction loss $\mathcal{L}_{\text{recon}}$. \textcolor{black}{The adversarial reconstruction loss component $\mathcal{L}_{adv\_recon}$ in this joint loss function is specifically designed to achieve the adversarial robustness objective.} Since the features of \(\bm{x}_n\) are relatively easy to learn while adversarial features are challenging to recover, we set $\alpha$ to 0.25. The \(\ell_1\) loss computes the absolute error between the model output and each element of the actual sample. By optimizing this loss, the Def-Transformer model can generate feature maps close to those of clean samples, effectively eliminating adversarial perturbation.

Regarding the choice of reconstruction loss function, the $\ell_1$ loss is preferred over the more commonly used $\ell_2$ loss for the following reasons: the $\ell_1$ loss is less sensitive to outliers, which prevents the model from overfitting to non-structural perturbations introduced by adversarial attacks. A detailed ablation analysis on the loss type and the selection of $\alpha$ is provided in Section~\ref{sec:ablation-exp}.

\subsection{\textcolor{black}{Theoretical Analysis via the Information Bottleneck}}
\textcolor{black}{The defense mechanism of T-ADD can be interpreted by the Information Bottleneck (IB) theory~\cite{tishby2015deeplearninginformationbottleneck}, which offers a information-theoretic framework for understanding how neural networks extract task-relevant features while suppressing irrelevant noise.}

\textcolor{black}{IB theory posits that a deep learning model learns an intermediate representation $Z$ that compresses the input random variable $X$ while preserving maxima mutual information about the target variable $Y$. This trade-off is typically formulated as the following Lagrangian objective~\cite{2020iblag}:
\begin{equation}
    \underset{{p(Z|X)}}{\max}\ I(Z; Y) - \beta I(X; Z)
\label{eq:ib}
\end{equation}
where $I(\cdot\,;\cdot)$ denotes mutual information and $\beta$ is a Lagrange multiplier that controls the degree of compression. In the context of T-ADD, the Def-Transformer with self-attention mechanisms is designed to filter out adversarial perturbations. The input $X_{\text{adv}}$ to the Def-Transformer consists of a clean sample $X$ and an adversarial perturbation $\delta$, i.e., $X_{\text{adv}} = X_{n} + \delta$. The output of the Def-Transformer is the reconstructed sample $X^{\text{re}}$, which is trained to approximate $X_{n}$. }
\textcolor{black}{By viewing this process through the IB theory, we aim to learn a representation $X^{\text{re}}$ that preserves relevant information about $X_{n}$ while discarding nuisance components such as adversarial perturbation. Substituting the relevant random variables into Eq.~\eqref{eq:ib}, we obtain the following objective:
\begin{equation}
    \underset{p(X^{\text{re}}|X_{\text{adv}})}{\max}\ I(X^{\text{re}}; X_{n}) - \beta I(X_{\text{adv}}; X^{\text{re}}).
    \label{eq:tadd_ib}
\end{equation}
This formulation captures the core goal of T-ADD: to produce a reconstruction $X^{\text{re}}$ that retains as much mutual information as possible with the clean sample $X_{n}$, while minimizing dependence on the adversarial sample $X_{\text{adv}}$, thereby suppressing the influence of adversarial perturbations. In T-ADD, this objective is optimized via a joint reconstruction loss and the global modeling capability of the self-attention mechanism, ensuring faithful signal recovery.}

\section{Simulations}
\subsection{Source Data Collection}\label{sec:dataset-setup}
We conducted simulations using MATLAB. We considered a scenario where BPSK-modulated signals were emitted onto an 8-element uniform linear array. The roll-off factor of the raised cosine filter for pulse shaping was set to 0.5. Subsequently, 1024 points of the signal are sampled and Additive White Gaussian Noise (AWGN) is added. Finally, the samples are obtained after power normalization.

For single signal source training, the SNR was uniformly distributed within the range of \(-14\) dB to \(10\) dB in steps of 2dB, with incident angles ranging from \(-90^\circ\) to $90^\circ$. At each SNR, ten samples per DOA were generated, culminating in a dataset of $23,530$ samples. For dual-source training, two DOA were chosen at random within the range of $-60^\circ$ to $60^\circ$. This selection process yielded a total of $7,260$ distinct DOA pairings. For each SNR level, every combination of DOAs produced two samples, ensuring that the SNR remained consistent with the configuration used in the single-source scenario.

The size of the validation set is half the number of samples used in training. It is worth emphasizing that the proposed method is trained and evaluated exclusively with PGD-based adversarial samples. MIM attack samples are excluded from both the training and validation phases to better simulate real-world deployment scenarios, where the nature of adversarial attacks is dynamic and typically unknown a priori.

The steps of attack were constrained to 5 with hyper-parameters tuned to a step size of 0.02 and a perturbation bound of 0.2. The SIR of the perturbation is maintained at 10 dB.

\subsection{Training Settings and Performance Metrics}
During the training process the defense model, the Adam optimizer was utilized, with the initial learning rate set to $1 \times 10^{-3}$. For single-source signal training, the rate was halved every five epochs. In dual-source case, it was reduced by half at the fifth epoch, then decreased by 50\% every two epochs. Training lasted for 30 epochs, with a batch size of 64. It is worth noting that adversarial samples are generated using only the PGD attack method during training.

To comprehensively evaluate the performance of the proposed method, several state-of-the-art defense methods are selected for comparison, including Adversarial Training (AT)~\cite{madry2019deeplearningmodelsresistant}, Randomized Smoothing (RS)~\cite{pmlr2019-cohenrandomsmooth}, and Robust Feature Inference (RFI)~\cite{singh2024robustfeatureinferencetesttime}.

The RMSE metric was utilized to assess the defense performance, which is defined as:
\begin{equation}
\text{RMSE} = \sqrt{\frac{1}{BL} \sum_{b=1}^B \sum_{l=1}^L \left( {\theta}_l^{(b)} - \hat{\theta}_l^{(b)} \right)^2 }, 
\end{equation}
where $B$ signifies the aggregate count of test samples, $L$ represents  the number of sources,  ${\theta}_l^{(b)}$  represents the DOA  for the $l$-th signal source in the $b$-th sample and $\hat{\theta}_l^{(b)}$ denotes the estimated DOA. RMSE reflects the overall estimation accuracy of the model, with lower values indicating better proximity to the ground truth.

To further assess the model’s performance on both clean and adversarial samples, we introduce an additional metric—accuracy. Due to the continuous nature of real-world DOAs, which are typically off-grid, traditional adversarial metric $\mathbb{I}$ are not directly applicable. Therefore, accuracy is formulated as:
\begin{equation}
    \text{Accuracy} = \frac{1}{B} \sum_{b=1}^B \bm{\Phi} \left( \left| {\bm{f}(\bm{x}^{(b)})} - \textcolor{black}{{y}}^{(b)} \right| \leq \zeta \right),
\end{equation}
where \textcolor{black}{${y}^{(b)}$} denotes the ground truth label of the $b$-th sample, $\bm{f}(\bm{x}^{(b)})$ is the corresponding prediction, and $\zeta > 0$ is a predefined threshold indicating the maximum tolerable angular deviation. The function $\bm{\Phi}(\cdot)$ returns 1 if the prediction error is within the acceptable range and 0 otherwise.
 
To quantify the model's prediction confidence across different DOA task types, we define task-specific confidence scores. For the single-source scenario, the model's output logits $c(\bm{x})$ are normalized via the softmax function. The confidence score $\text{Conf}_{\text{1}}$ is defined as the maximum probability over all classes:
  \begin{equation}
  \text{Conf}_{\text{1}} = \max\left( \text{Softmax}\left( c(\bm{x}) \right) \right),
  \end{equation}
where $\max(\cdot)$ denotes the highest value in the resulting probability vector.

For the dual-source scenario, the model utilizes the sigmoid function, generating independent class probabilities. The confidence score $\text{Conf}_{\text{2}}$ is defined as the average of the two highest probabilities in the output vector:

  \begin{equation}
  \text{Conf}_{\text{2}} = \frac{1}{2} \left( \max\nolimits_1\left( \Psi(c(\bm{x})) \right) + \max\nolimits_2\left( \Psi(c(\bm{x})) \right) \right),
  \end{equation}
where $\Psi(\cdot)$ denotes the sigmoid function, $\max\nolimits_1$ denotes the highest value, and $\max\nolimits_2$ denotes the second-highest value. To minimize discretization artifacts arising from rounding probabilities typically displayed with two decimal places, all confidence scores are expressed as percentages.

When employing RS with Monte Carlo sampling, the confidence score $\text{Conf}_{\text{RS}}$ is measured as the empirical frequency of the most frequent predicted class relative to the total number of trials $G_{\text{sum}}$. For the single-source case:

\begin{equation}
    \text{Conf}_{\text{RS}} = \frac{G}{G_{\text{sum}}},
\end{equation}
where $G$ represents the mode frequency. For the dual-source scenario, $\text{Conf}_{\text{RS}}$ averages the frequencies of the two most frequent predictions by summing them and dividing by $G_{\text{sum}}$ to estimate the confidence score. These scores are also reported as percentages.

\begin{figure}[tbp]
\centering
\subfigure[]{%
  \label{fig:subfig1:alpha_sig1_test_clean}
  \includegraphics[width=0.44425\linewidth]{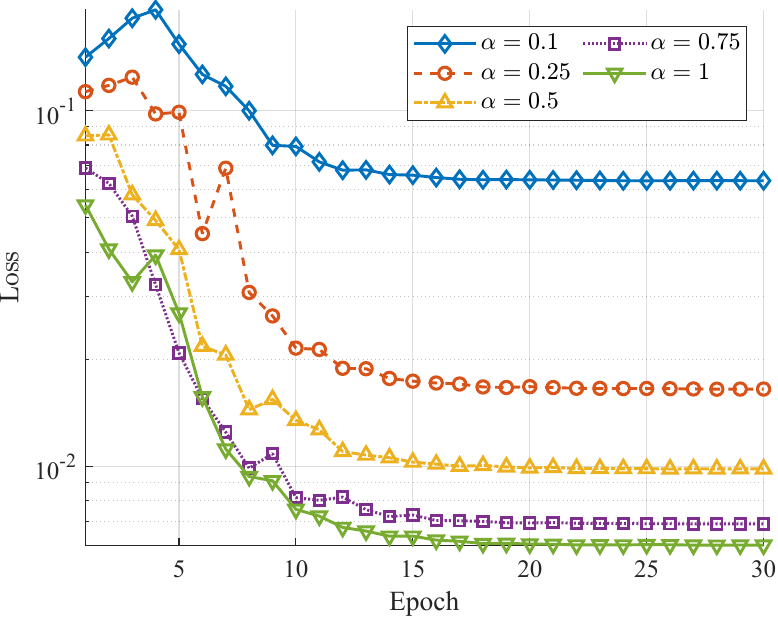}%
}
\subfigure[]{%
  \label{fig:subfig2:alpha_sig1_test_adv}
  \includegraphics[width=0.44425\linewidth]{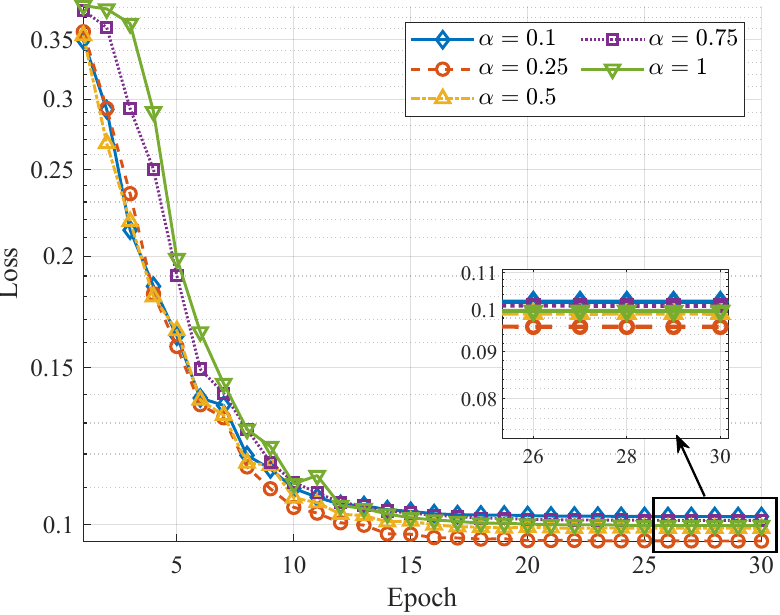}%
}

\subfigure[]{%
  \label{fig:subfig1:grid_clean}
  \includegraphics[width=0.44425\linewidth]{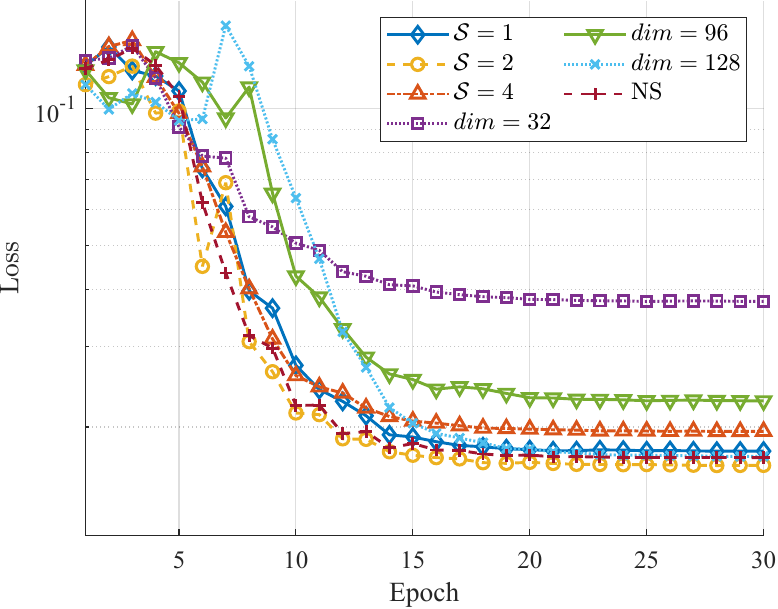}%
}
\subfigure[]{%
  \label{fig:subfig2:grid_adv}
  \includegraphics[width=0.44425\linewidth]{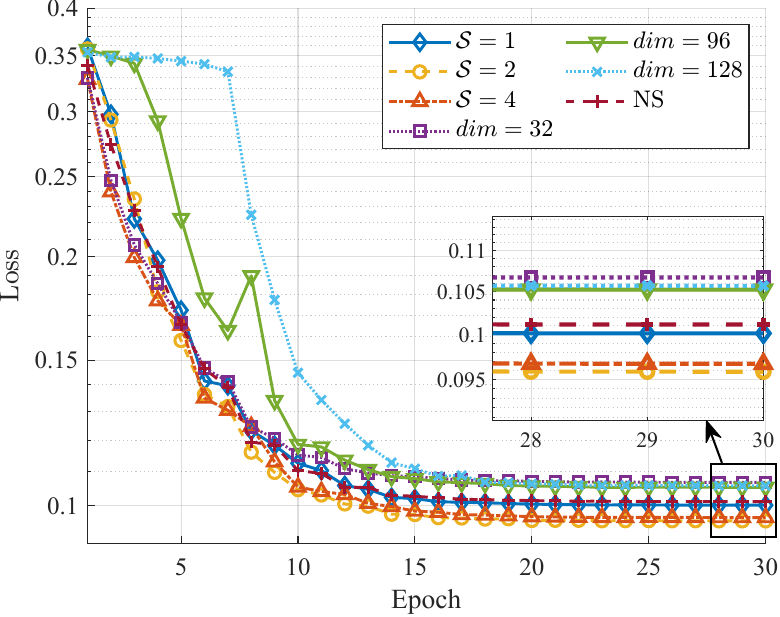}%
}

\caption{Losses of validating under different settings of parameter. (a) $L=1$, $\alpha$, $\mathcal{L}_{recon}$ (b) $L=1$, $\alpha$, $\mathcal{L}_{adv\_recon}$ (c) $L=1$, other parameters, $\mathcal{L}_{recon}$ (d) $L=1$, other parameters, $\mathcal{L}_{adv\_recon}$.}
\label{fig:alpha-search}
\end{figure}

\subsection{Ablation studies}\label{sec:ablation-exp}
In this experiment, we investigate the impact of \textcolor{black}{different hyperparameters} on model performance. \textcolor{black}{Robustness in the ablation experiments is evaluated from the perspective of reconstruction loss. A smaller adversarial reconstruction loss $\mathcal{L}_{adv\_recon}$, indicates that the reconstructed sample remains closer to the clean sample under adversarial attack, thereby reflecting stronger robustness~\cite{stutz2021robustloss}.}

We first examine the influence of the weight parameter $\alpha$ in the loss function. A grid search is conducted over $\alpha \in \{0.1, 0.25, 0.5, 0.75, 1.0\}$. Subsequently, we perform additional grid searches for the downsampling factor $\mathcal{S}$ and the embedding dimension $dim$, where $\mathcal{S} \in \{1,2,4\}$ and $dim \in \{32,64, 96, 128\}$. Furthermore, a variant of the model without the self-attention module in Swin V2 is also included. The training set follows the same configuration described in Section~\ref{sec:dataset-setup}, while the validation set size is set to half of the training set. \textcolor{black}{Fig.~\ref{fig:alpha-search}} shows the reconstruction loss and adversarial reconstruction loss on the validation set across different training epochs \textcolor{black}{under hyperparameter settings, where ``NS'' denotes model without self-attention module in Swin V2}. 

Experimental results indicate that when $\alpha = 0.25$, the model achieves the lowest adversarial reconstruction loss. As $\alpha$ increases, the reconstruction loss on clean samples gradually decreases, suggesting that the model’s ability to fit clean data is positively correlated with a larger $\alpha$. Although $\alpha = 0.25$ is not optimal on clean samples, in practical applications, a trade-off between robustness and clean sample performance must be considered. In real-world DOA estimation scenarios, erroneous predictions caused by adversarial attacks can be catastrophic. Hence, we select $\alpha = 0.25$ as the optimal hyperparameter setting for the subsequent experiments.

\textcolor{black}{To further analyze robustness}, we evaluated the performance of models trained using the Mean Squared Error (MSE) loss function. The model trained with MSE achieved a reconstruction loss of 0.069 and an adversarial reconstruction loss of 0.142 on the validation set, both significantly higher than the corresponding values (0.016 and 0.096) achieved by the model trained with the \(\ell_1\) loss. \textcolor{black}{Since a smaller $\mathcal{L}_{adv\_recon}$ implies stronger robustness, this result confirms that the \(\ell_1\) loss leads to better resistance against adversarial attacks.}

Next, we isolate the effects of varying $\mathcal{S}$ and $dim$ while keeping other settings fixed. For the experiments varying $\mathcal{S}$, $dim$ is set to 64. For experiments varying $dim$, $\mathcal{S}$ is set to 2. Results show that the configuration $\mathcal{S}=2$ and $dim=64$ leads to the lowest adversarial reconstruction loss $\mathcal{L}_{adv\_recon}$ and clean reconstruction loss $\mathcal{L}_{recon}$. Additionally, removing the self-attention mechanism from Swin V2 significantly degrades model performance, underscoring its importance in enhancing robustness.

\textcolor{black}{The loss curves in Fig.~\ref{fig:alpha-search} exhibit no signs of robust overfitting—unlike adversarial training, where the robust validation loss eventually increases due to over-specialization~\cite{wang2019convergence}. In contrast, T-ADD effectively suppresses adversarial perturbations, as evidenced by the monotonic convergence of both $\mathcal{L}_{recon}$ and $\mathcal{L}_{adv\_recon}$.}

\begin{figure*}[tbp]
\centering
\subfigure[]{\label{fig:pgd_sig1_xsnr_rmse}
\includegraphics[width=0.23\linewidth]{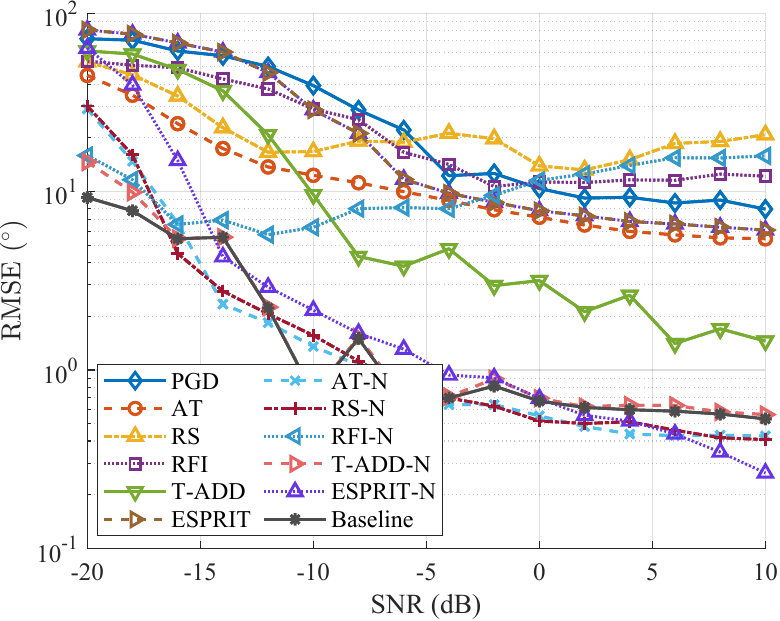}}
\subfigure[]{\label{fig:mim_sig1_xsnr_rmse}
\includegraphics[width=0.23\linewidth]{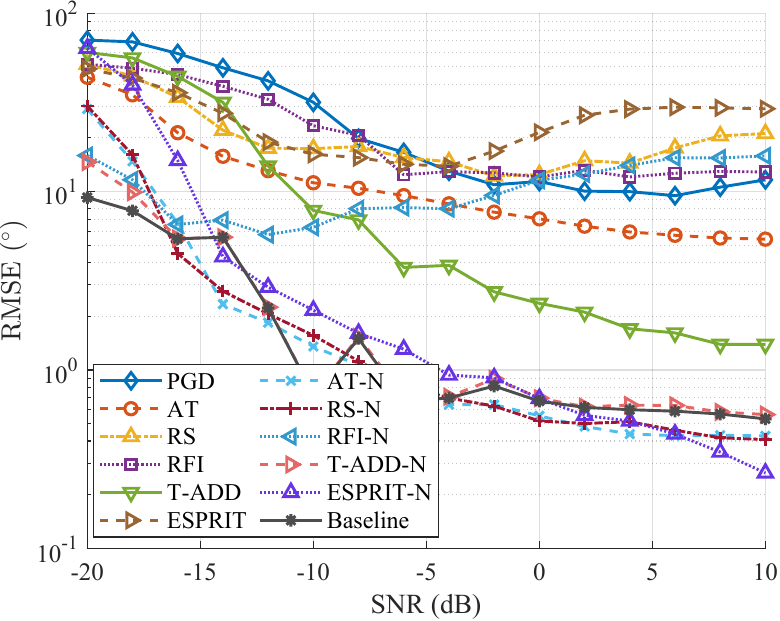}}
\subfigure[]{\label{fig:pgd_sig2_xsnr_rmse}
\includegraphics[width=0.23\linewidth]{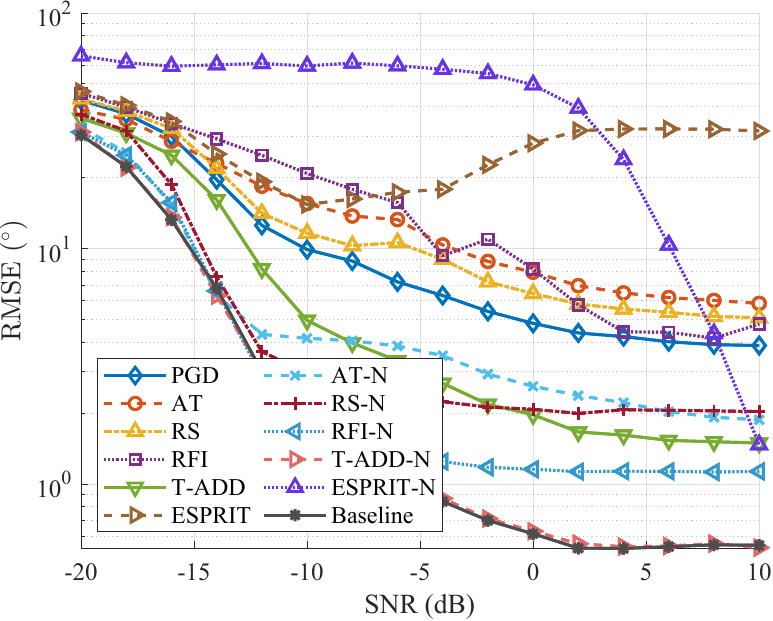}}
\subfigure[]{\label{fig:mim_sig2_xsnr_rmse}
\includegraphics[width=0.23\linewidth]{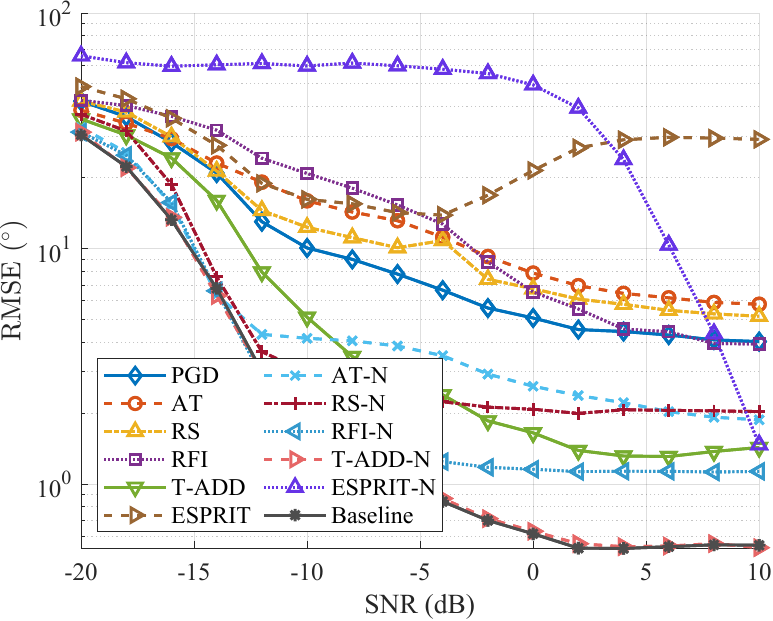}}

\subfigure[]{\label{fig:pgd_sig1_xsnr_acc}
\includegraphics[width=0.23\linewidth]{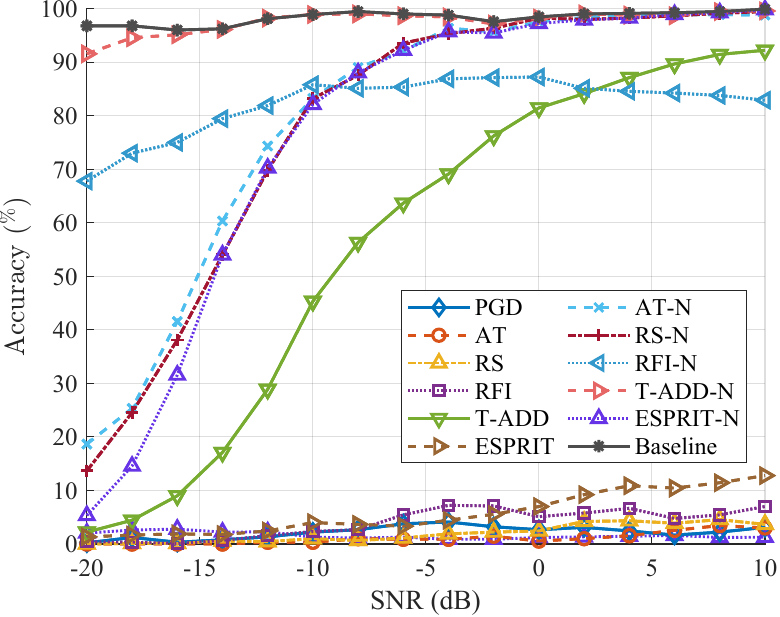}}
\subfigure[]{\label{fig:mim_sig1_xsnr_acc}
\includegraphics[width=0.23\linewidth]{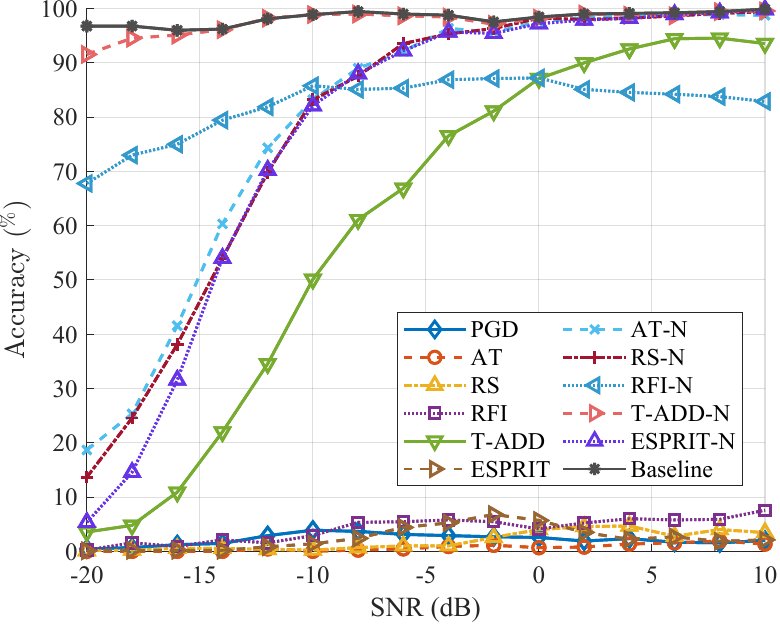}}
\subfigure[]{\label{fig:pgd_sig2_xsnr_acc}
\includegraphics[width=0.23\linewidth]{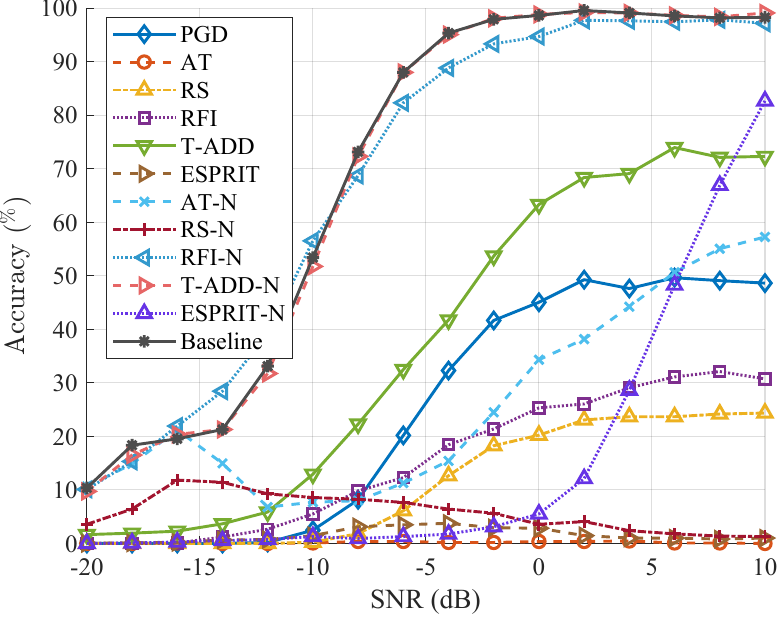}}
\subfigure[]{\label{fig:mim_sig2_xsnr_acc}
\includegraphics[width=0.23\linewidth]{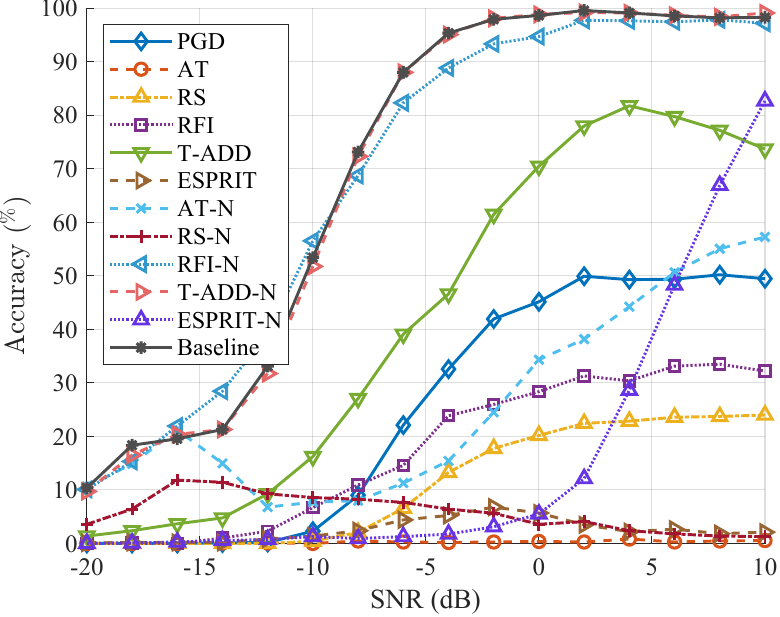}}

\subfigure[]{\label{fig:pgd_sig1_xsnr_conf}
\includegraphics[width=0.23\linewidth]{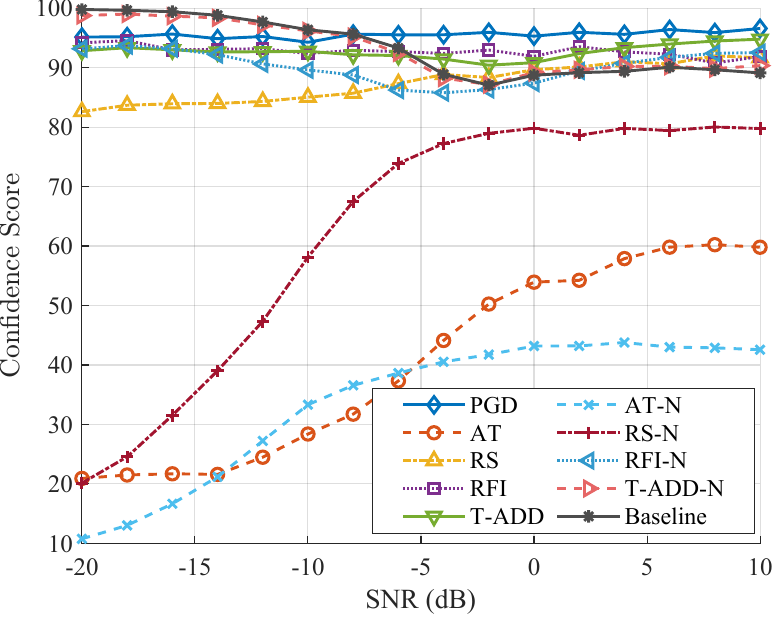}}
\subfigure[]{\label{fig:mim_sig1_xsnr_conf}
\includegraphics[width=0.23\linewidth]{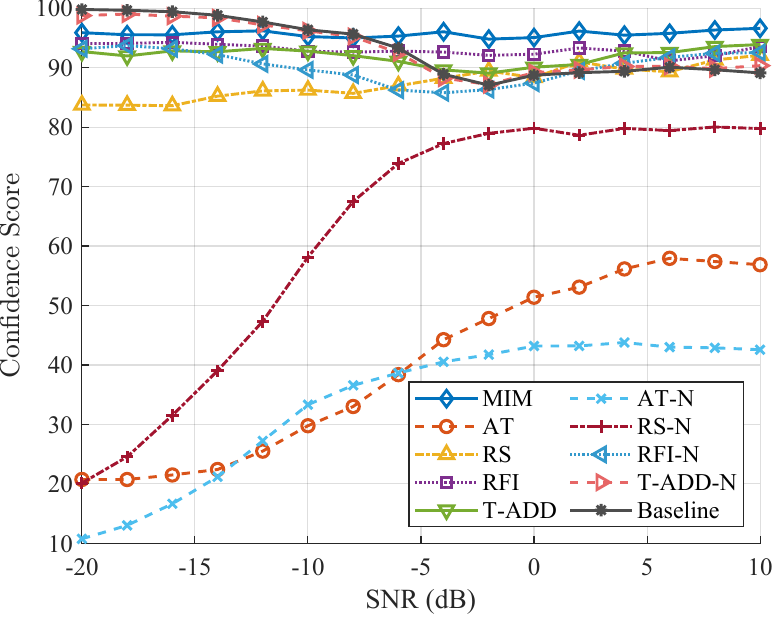}}
\subfigure[]{\label{fig:pgd_sig2_xsnr_conf}
\includegraphics[width=0.23\linewidth]{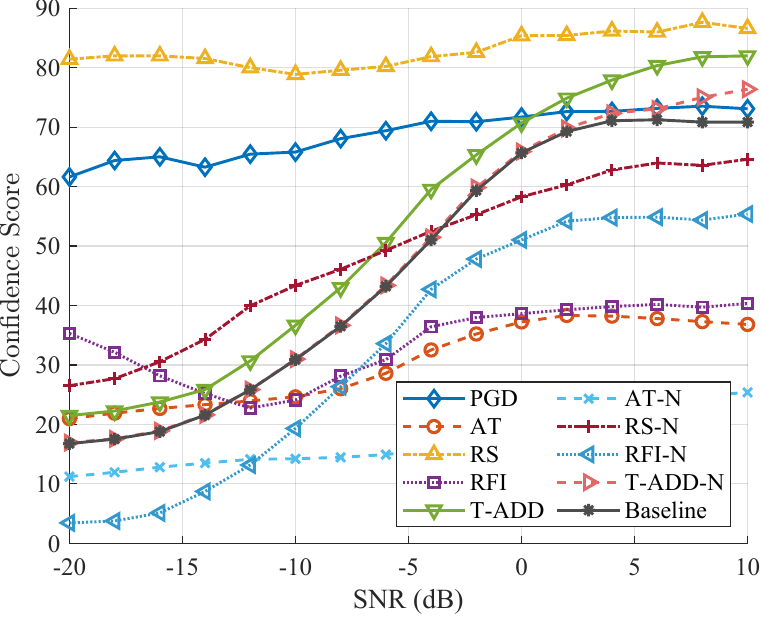}}
\subfigure[]{\label{fig:mim_sig2_xsnr_conf}
\includegraphics[width=0.23\linewidth]{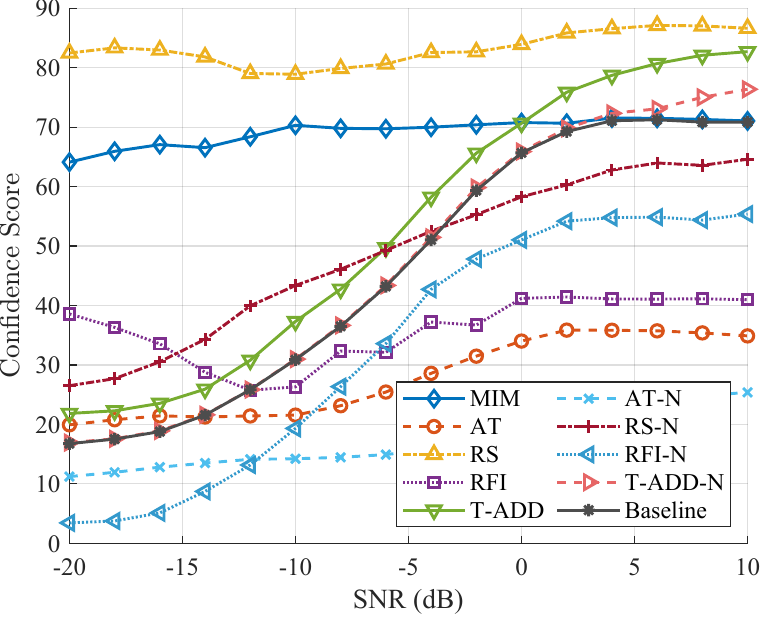}}
\centering
\caption{Comparison of different defense methods under PGD and MIM attacks across varying SNRs. Comparison of different defense methods under PGD and MIM attacks across varying SIRs. (a) RMSE under PGD attack with $L=1$, (b) RMSE under MIM with $L=1$, (c) RMSE under PGD with $L=2$, (d) RMSE under MIM with $L=2$, (e) Accuracy under PGD with $L=1$, (f) Accuracy under MIM with $L=1$, (g) Accuracy under PGD with $L=2$, (h) Accuracy under MIM with $L=2$, (i) Confidence score under PGD with $L=1$, (j) Confidence score under MIM with $L=1$, (k) Confidence score under PGD with $L=2$ and (l) Confidence score under MIM with $L=2$.}\label{fig-snr}
\end{figure*}
\subsection{Simulation Results}
\subsubsection{Comparison under Different SNRs}\label{section-exp-snr}

\begin{figure*}[tbp]
\centering
\subfigure[]{\label{fig:pgd_sig1_xkp_rmse}
\includegraphics[width=0.23\linewidth]{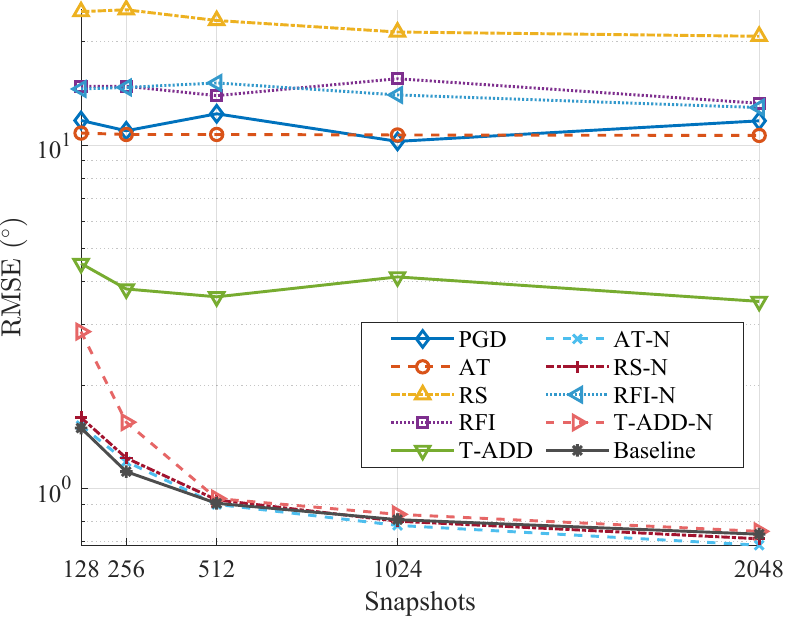}}
\subfigure[]{\label{fig:mim_sig1_xkp_rmse}
\includegraphics[width=0.23\linewidth]{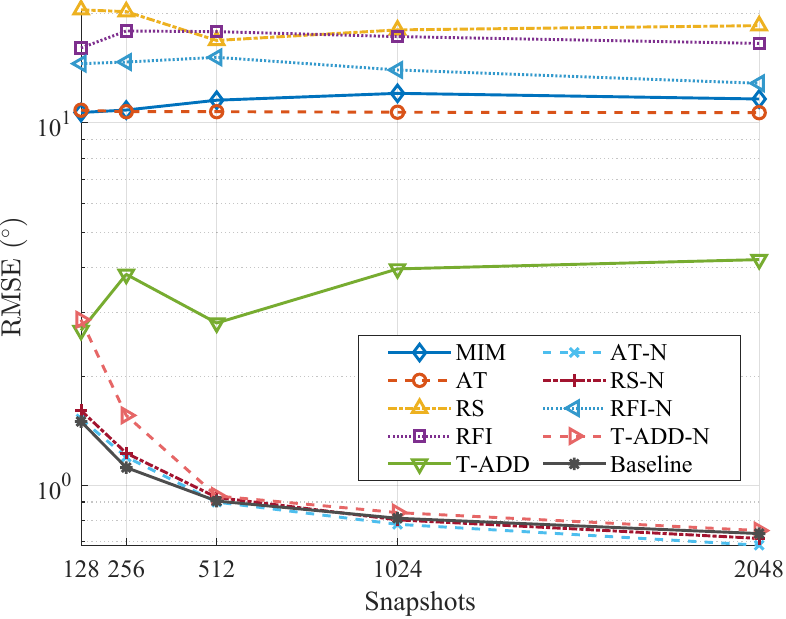}}
\subfigure[]{\label{fig:pgd_sig2_xkp_rmse}
\includegraphics[width=0.23\linewidth]{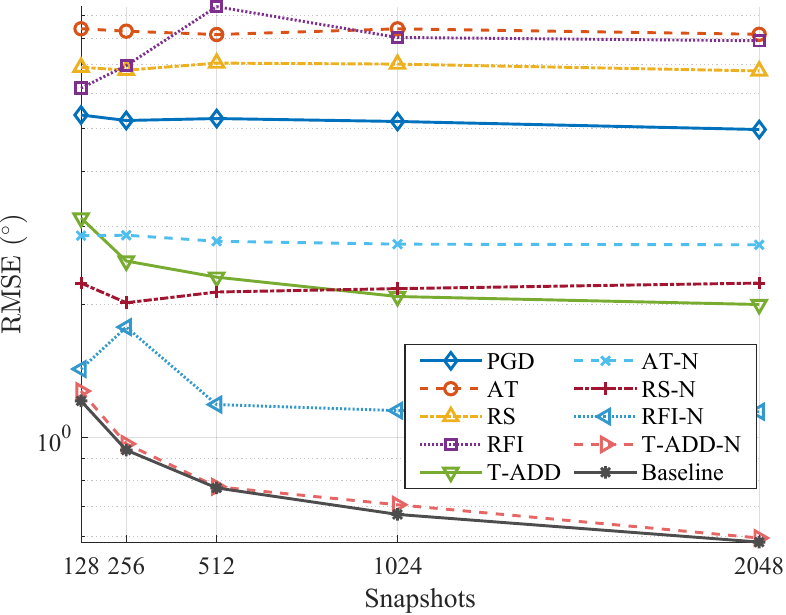}}
\subfigure[]{\label{fig:mim_sig2_xkp_rmse}
\includegraphics[width=0.23\linewidth]{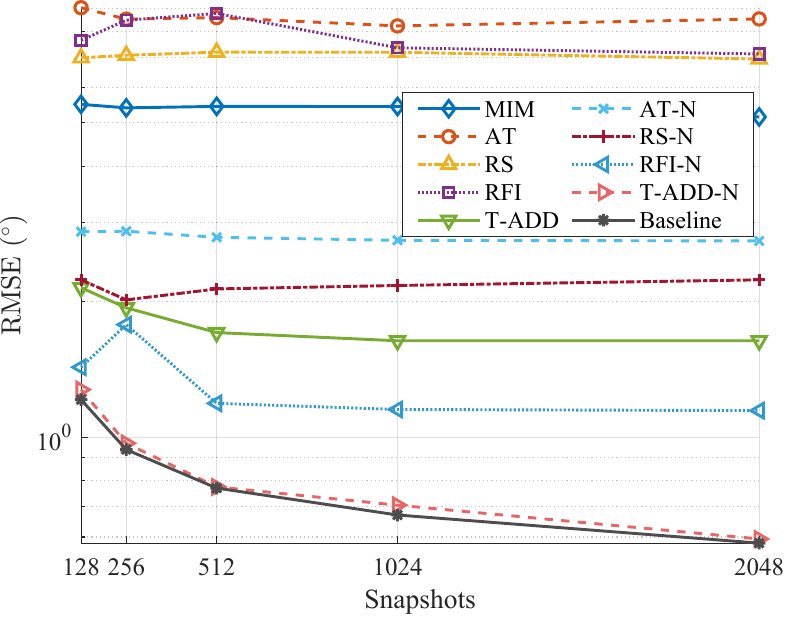}}

\subfigure[]{\label{fig:pgd_sig1_xkp_acc}
\includegraphics[width=0.23\linewidth]{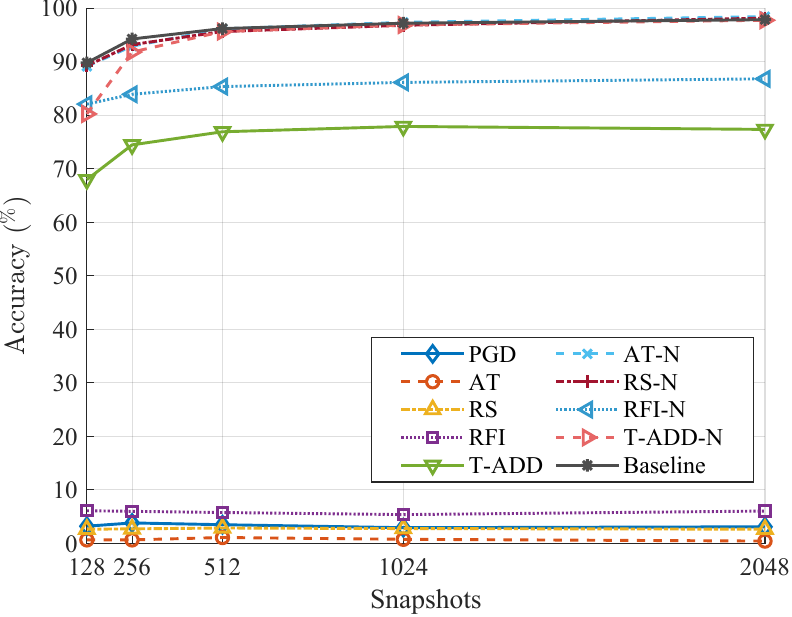}}
\subfigure[]{\label{fig:mim_sig1_xkp_acc}
\includegraphics[width=0.23\linewidth]{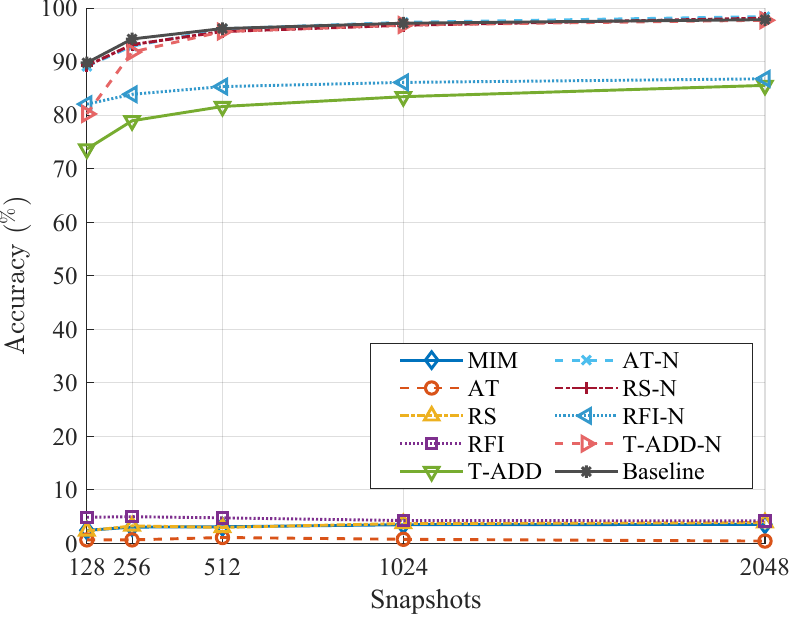}}
\subfigure[]{\label{fig:pgd_sig2_xkp_acc}
\includegraphics[width=0.23\linewidth]{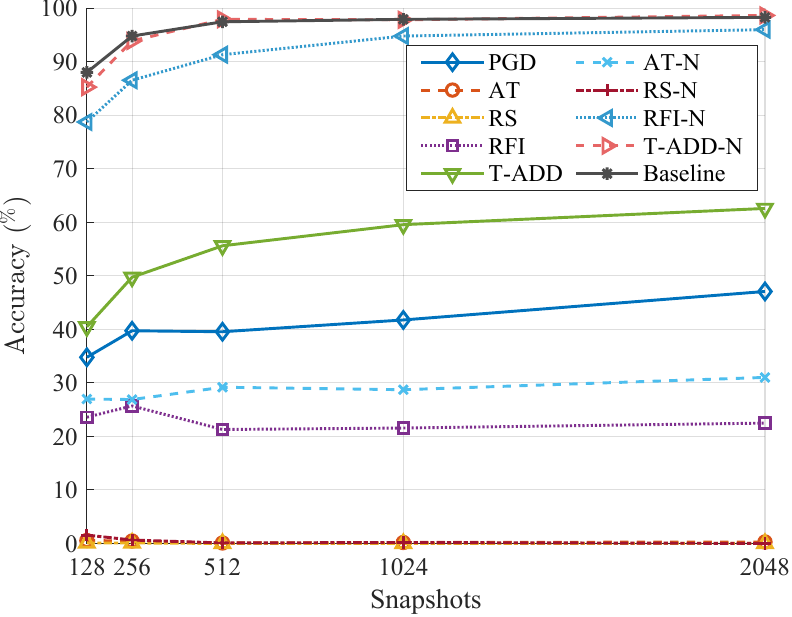}}
\subfigure[]{\label{fig:mim_sig2_xkp_acc}
\includegraphics[width=0.23\linewidth]{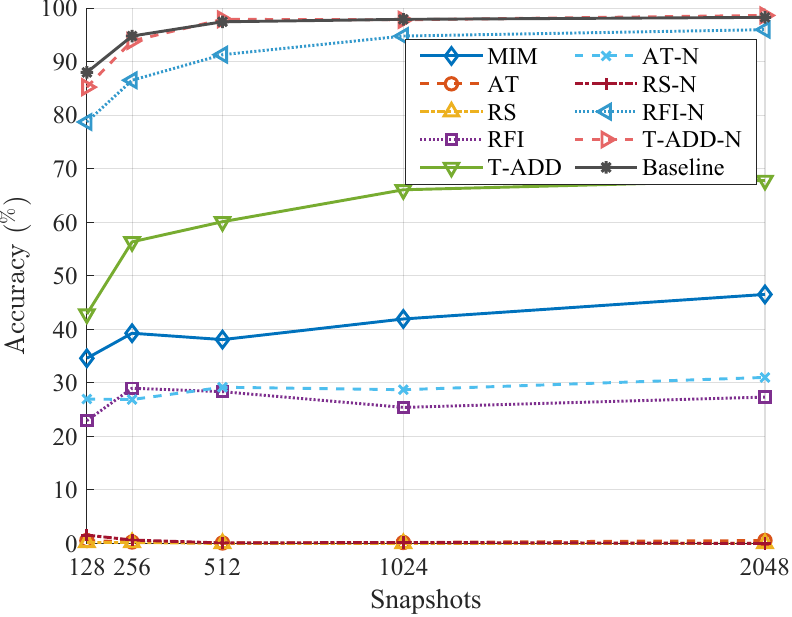}}

\subfigure[]{\label{fig:pgd_sig1_xkp_conf}
\includegraphics[width=0.23\linewidth]{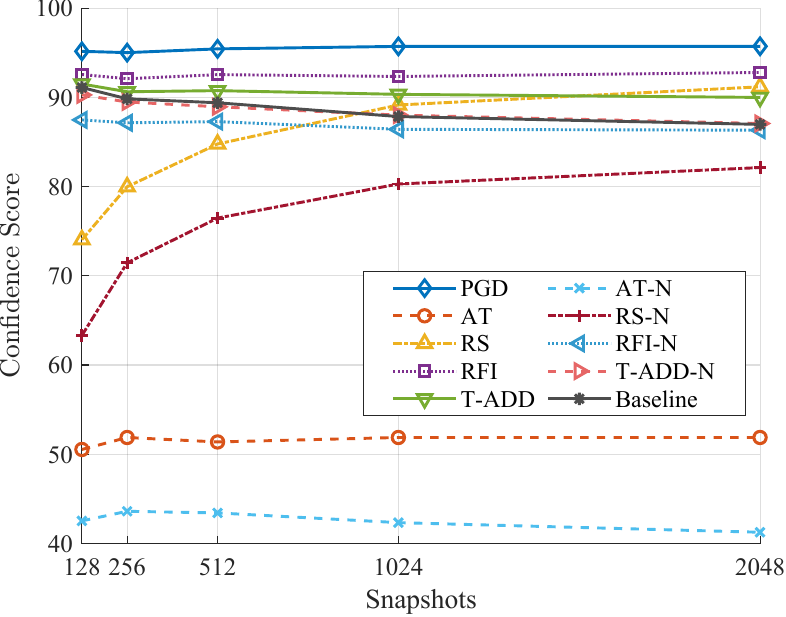}}
\subfigure[]{\label{fig:mim_sig1_xkp_conf}
\includegraphics[width=0.23\linewidth]{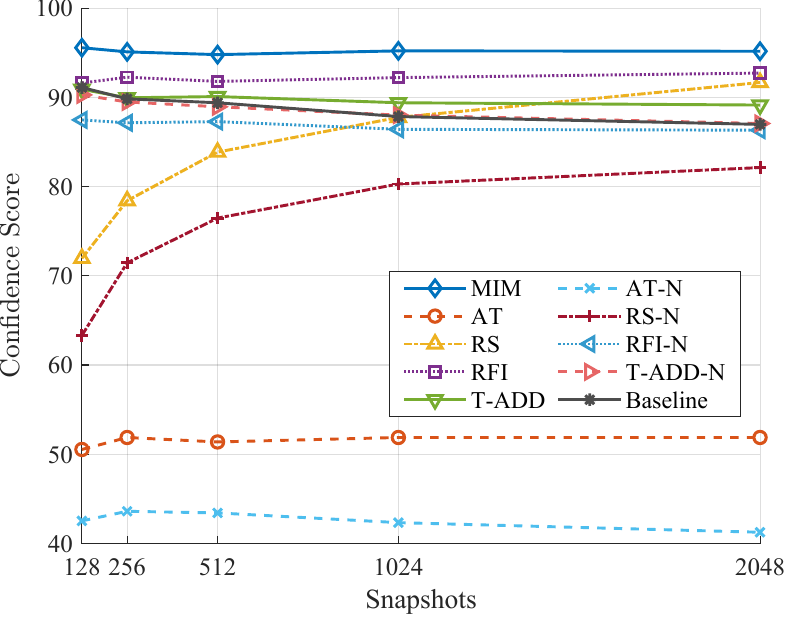}}
\subfigure[]{\label{fig:pgd_sig2_xkp_conf}
\includegraphics[width=0.23\linewidth]{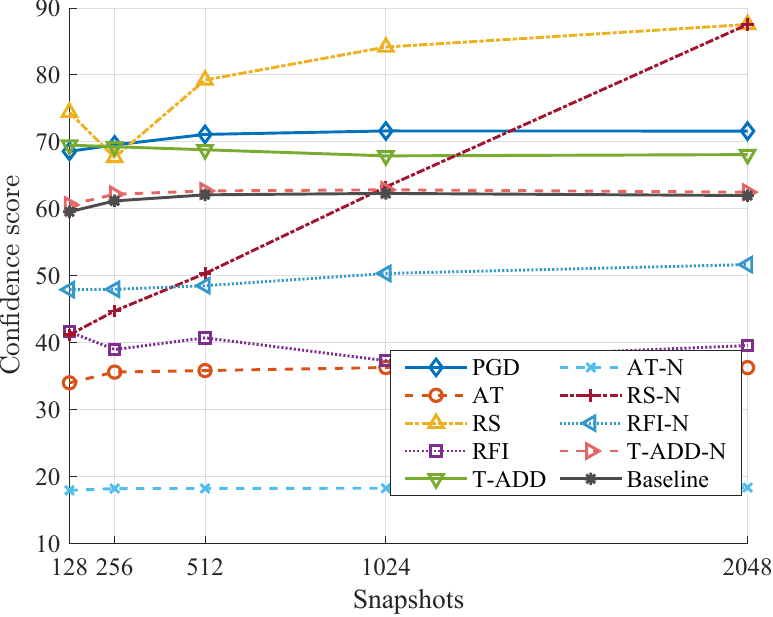}}
\subfigure[]{\label{fig:mim_sig2_xkp_conf}
\includegraphics[width=0.23\linewidth]{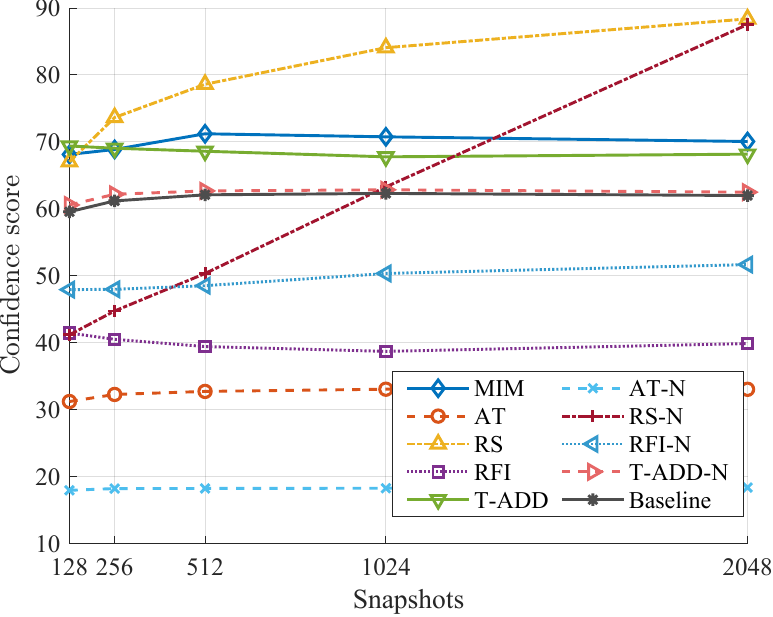}}
\centering
\caption{Comparison of different defense methods under PGD and MIM attacks across varying number of snapshots. Comparison of different defense methods under PGD and MIM attacks across varying SIRs. (a) RMSE under PGD attack with  $L=1$, (b) RMSE under MIM with $L=1$, (c) RMSE under PGD with $L=2$, (d) RMSE under MIM with $L=2$, (e) Accuracy under PGD with $L=1$, (f) Accuracy under MIM with $L=1$, (g) Accuracy under PGD with $L=2$, (h) Accuracy under MIM with $L=2$, (i) Confidence score under PGD with $L=1$, (j) Confidence score under MIM with $L=1$, (k) Confidence score under PGD with $L=2$ and (l) Confidence score under MIM with $L=2$.}
\label{fig-snapshot}
\end{figure*}
Firstly, we focus on the defense performance under different SNRs for signal sources with $L=1,2$. For $L=1$, the angles considered range from $-89.5^\circ$ to $89.5^\circ$ with a resolution of $1^\circ$, while for $L=2$, the angles range from $-55.5^\circ$ to $55.5^\circ$ with a resolution of $1^\circ$ and an angular separation of $3^\circ$ between each signal. This setup corresponds to scenarios where all angles are outside of the training dataset. For \( L = 1 \), we generated 5 clean samples for each DOA, and for \( L = 2 \), 10 clean samples were generated for each pair of DOA. Adversarial samples corresponding to these clean samples were then generated using the attack method. To comprehensively evaluate robustness under realistic channel conditions, we vary the SNR from \(-20\,\mathrm{dB}\) to \(10\,\mathrm{dB}\) during testing.
In the experiments, the steps of PGD and MIM attack is increased to 10, implementing a more intense attack to more effectively assess the robustness of the method. For evaluations of MIM, the decay factor is set to 1. It is important to note that all DOAs used in this test are entirely unseen during training.
Given that test DOAs are not included in the training set, larger estimation errors are observed. Thus, the threshold $\zeta$ for acceptable prediction deviation in accuracy evaluation is set to 2. These settings are consistently applied across all experiments presented below, unless otherwise stated.
The performance of RMSE under various SNRs with a number of snapshots $K=1024$ is illustrated in \textcolor{black}{Fig. \ref{fig-snr}}.  In \textcolor{black}{this} figure, ``Baseline'' indicates the baseline performance of the DOA estimation model when evaluated on clean samples, while ``-N'' represents the performance of a certain method on clean samples.

The experimental results demonstrate that the proposed adversarial defense method significantly enhances the model’s robustness compared to other approaches, while incurring only a minimal performance drop on clean samples. Notably, the proposed method not only effectively defends against PGD attacks, but also maintains strong robustness against MIM attacks, which are not observed during training. For comparison, we also include the traditional ESPRIT \cite{roy1989esprit} method. As shown in \textcolor{black}{Fig. \ref{fig-snr}}, its RMSE remains high under white-box adversarial attacks, which reflects its lack of robustness against these attacks.

Specifically, at an SNR of 6 dB with $L = 1$, the proposed method reduces the RMSE under PGD and MIM attacks from 8.62 and 9.48 to 1.41 and 1.61, achieving reductions of 83.6\% and 83.0\%, respectively. Notably, this substantial improvement in robustness is attained with minimal sacrifice in performance on clean inputs. For $L = 2$, across the SNR range of $-10$ dB to 10 dB, the maximum RMSE increase of T-ADD-N compared to the Baseline is only 0.03, consistently demonstrating the method's ability to effectively balance robustness and estimation accuracy.

In terms of accuracy, when $L=1$, the proposed method achieves over 87\% accuracy under both PGD and MIM attacks at an SNR of 4 dB. For $L = 2$, T-ADD improves the accuracy under MIM attacks from 49.27\% to 81.74\%. In contrast, the highest accuracy achieved by AT, RS, and RFI under MIM attacks in the $L = 2$ scenario does not exceed 33.49\%. On clean samples at 4 dB SNR, the accuracy of T-ADD differs from that of the baseline model by only 1.56\% and 1.83\% for $L = 1$ and $L = 2$, respectively.

Furthermore, Welch's t-test was conducted to statistically evaluate the performance differences between the T-ADD model and the Baseline method on clean samples. All p-values exceed the predefined significance level of \(\alpha = 0.05\). The statistical results which across a SNR range from \(-16\,\mathrm{dB}\) to \(10\,\mathrm{dB}\) show that when \(L=1\), the p-value for accuracy is no less than 0.08; and when \(L=2\), the corresponding p-value is no less than 0.09.  This statistical evidence indicates that the T-ADD model consistently maintains comparable accuracy to the Baseline method across different SNRs, with no statistically significant differences.

Regarding confidence scores on adversarial samples, we observe that the undefended model still outputs high confidence scores in the case of $L = 1$, despite being misled. This behavior is likely due to adversarial perturbations severely distorting the decision boundaries, causing the model to misclassify inputs into high confidence scores but incorrect classes outside the training distribution. In contrast, T-ADD maintains high confidence scores while achieving high accuracy on adversarial samples. Meanwhile, RS and RFI tend to produce high confidence scores even for incorrect predictions, further highlighting the effectiveness of our method. In the $L = 2$ case, both the confidence score and accuracy of AT and RFI degrade significantly under adversarial attacks, revealing their limitations in handling adversarial attacks during testing.

\subsubsection{Comparison under Different Number of Snapshots}

In this experimental setup, the angular range is consistent with that described in Section~\ref{section-exp-snr}. For each scenario, 10 clean samples are generated for each DOA, which are used as the validation set. A range of snapshot numbers $K \in \{128, 256, 512, 1024, 2048\}$ is considered to evaluate performance under a fixed SNR of $-1$ dB. Fig.~\ref{fig-snapshot} presents the RMSE, accuracy, and confidence scores of different methods under varying snapshot numbers.

The experimental results demonstrate that the proposed method consistently achieves strong robustness across different $K$, outperforming other defense methods. Moreover, the proposed method generalizes well to MIM attacks, which were not encountered during training.

Specifically, T-ADD significantly reduces the RMSE compared to AT, RFI, RS, and the undefended baseline across all values of $K$, while T-ADD-N maintains a low RMSE comparable to that of the Baseline. For example, in the challenging case of dual-source MIM attacks with $K = 128$, our method reduces the RMSE from 5.50 to 2.15—lower even than the clean sample RMSE achieved by AT and RS under the same condition.
In terms of accuracy, T-ADD exhibits a clear upward trend as the number of snapshots increases. Under MIM attacks with $L=1$, the accuracy of T-ADD improves from 73.67\% at $K = 128$ to 85.56\% at $K = 2048$. In contrast, other methods fail to show a consistent improvement with increasing $K$, indicating limited adaptability to increased snapshots and insufficient robustness to adversarial perturbations.

Regarding confidence scores, RS shows a gradual increase as $K$ increases, while the confidence scores of other methods remain relatively unchanged. Notably, T-ADD consistently produces confidence scores close to those of the Baseline across various $K$. This suggests that T-ADD effectively remaps adversarial samples back to the distribution of clean samples, maintaining high confidence scores even under adversarial attacks.


\subsubsection{Comparison under Different SIRs}
\begin{figure*}[tbp]
\centering
\subfigure[]{\label{fig:pgd_sig1_xsir_rmse}
\includegraphics[width=0.23\linewidth]{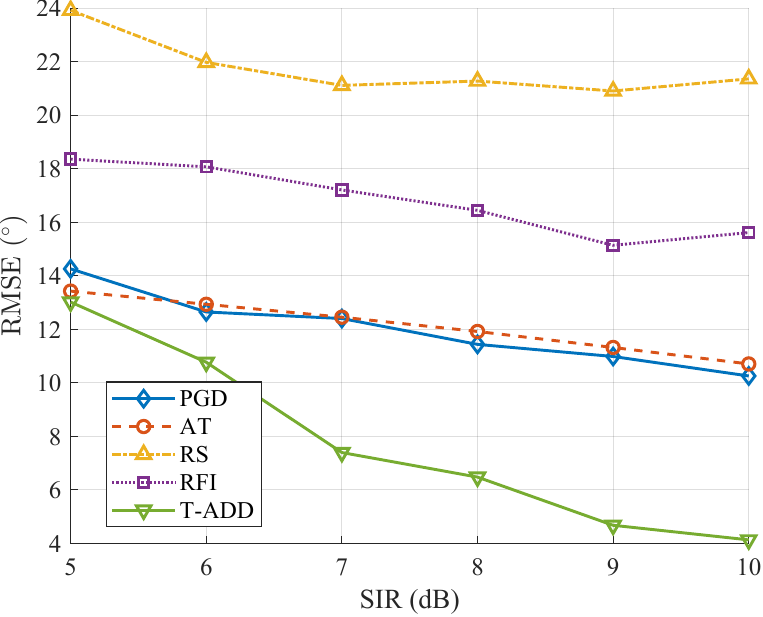}}
\subfigure[]{\label{fig:mim_sig1_xsir_rmse}
\includegraphics[width=0.23\linewidth]{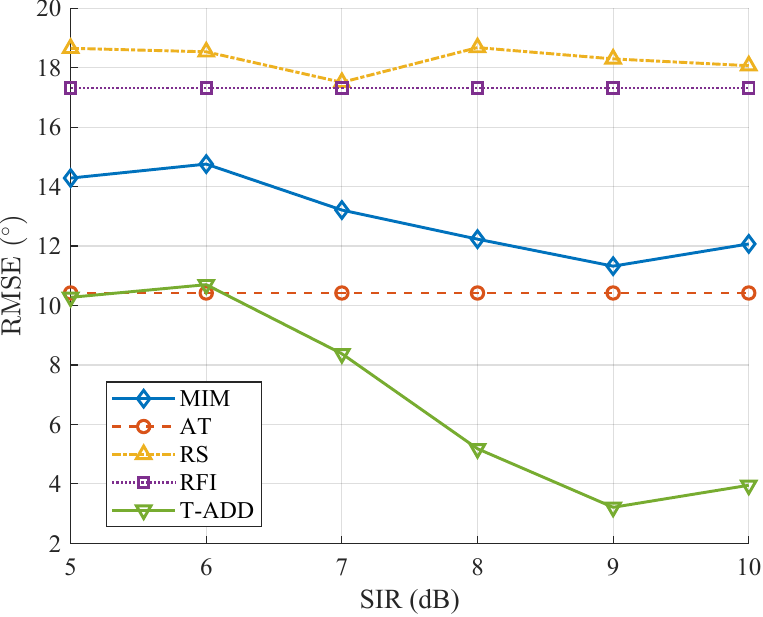}}
\subfigure[]{\label{fig:pgd_sig2_xsir_rmse}
\includegraphics[width=0.23\linewidth]{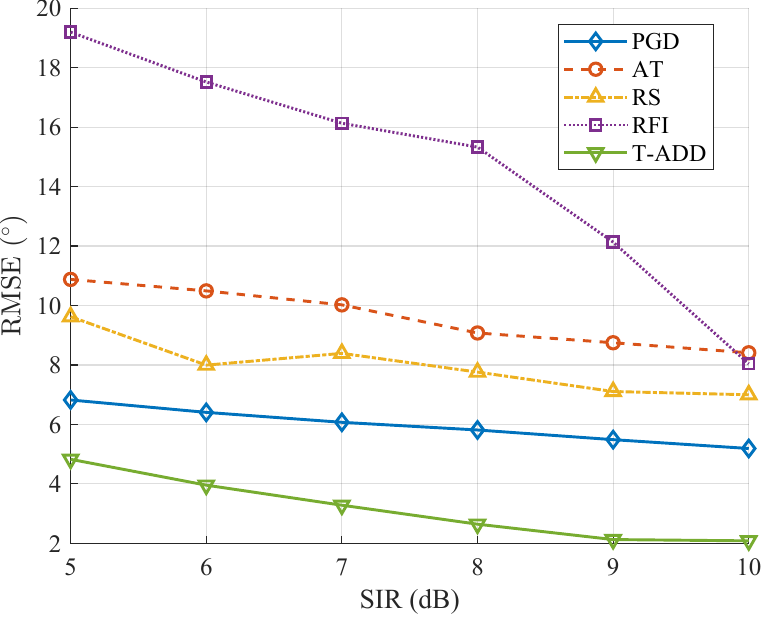}}
\subfigure[]{\label{fig:mim_sig2_xsir_rmse}
\includegraphics[width=0.23\linewidth]{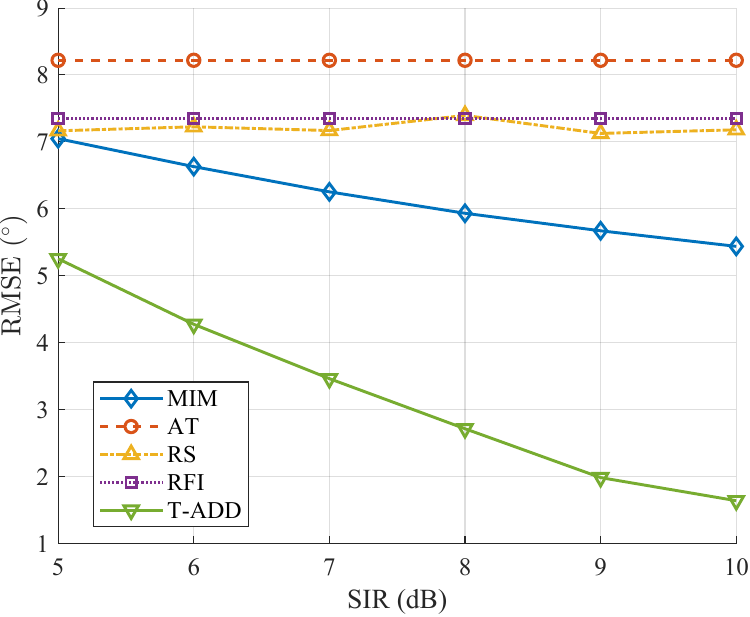}}

\subfigure[]{\label{fig:pgd_sig1_xsir_acc}
\includegraphics[width=0.23\linewidth]{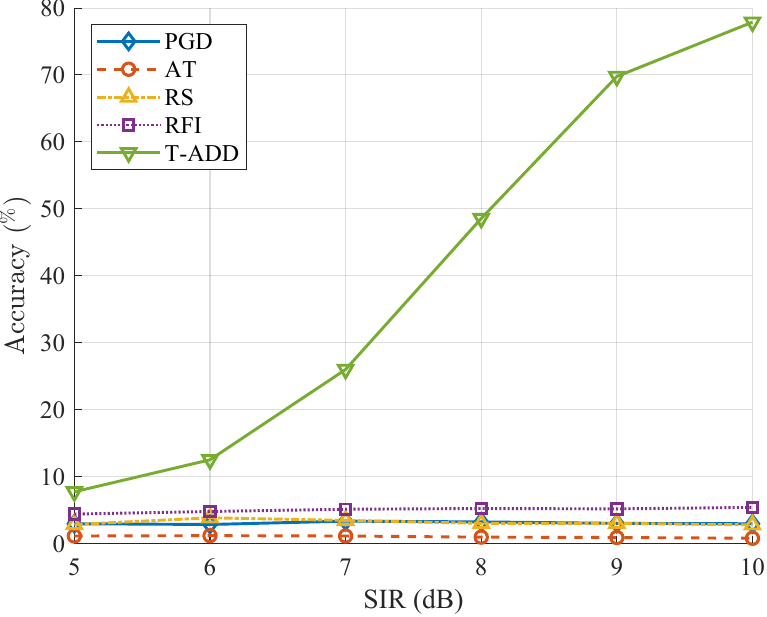}}
\subfigure[]{\label{fig:mim_sig1_xsir_acc}
\includegraphics[width=0.23\linewidth]{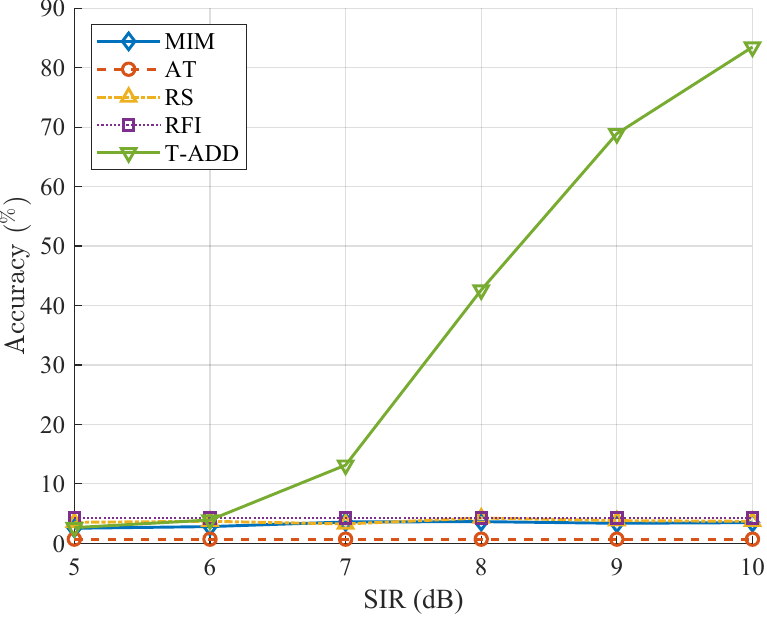}}
\subfigure[]{\label{fig:pgd_sig2_xsir_acc}
\includegraphics[width=0.23\linewidth]{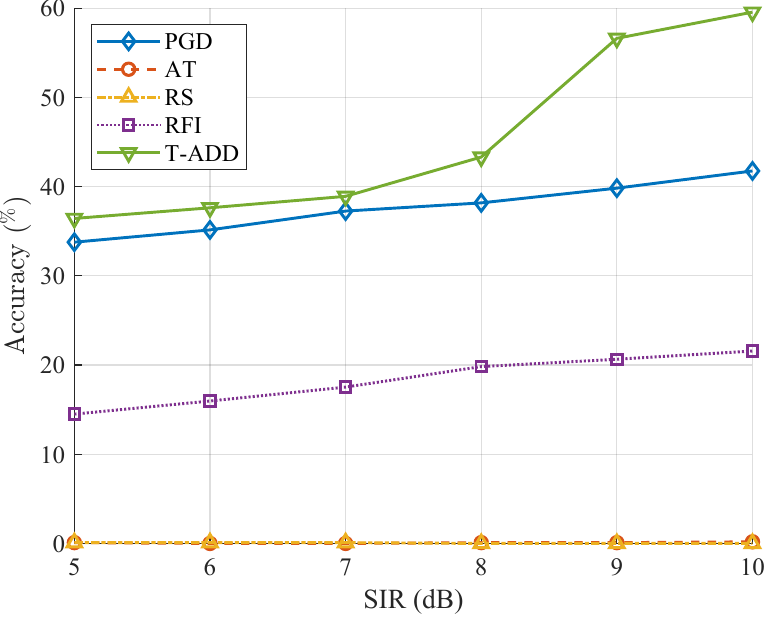}}
\subfigure[]{\label{fig:mim_sig2_xsir_acc}
\includegraphics[width=0.23\linewidth]{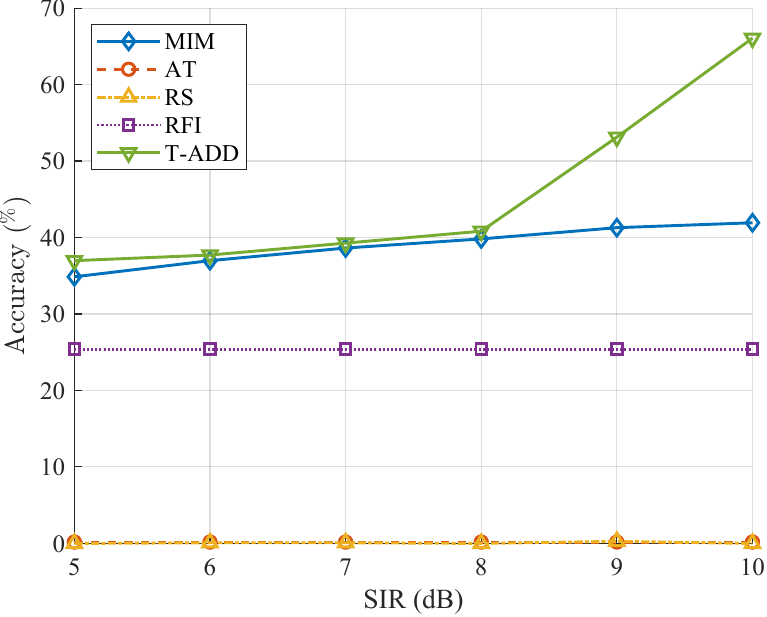}}

\subfigure[]{\label{fig:pgd_sig1_xsir_conf}
\includegraphics[width=0.23\linewidth]{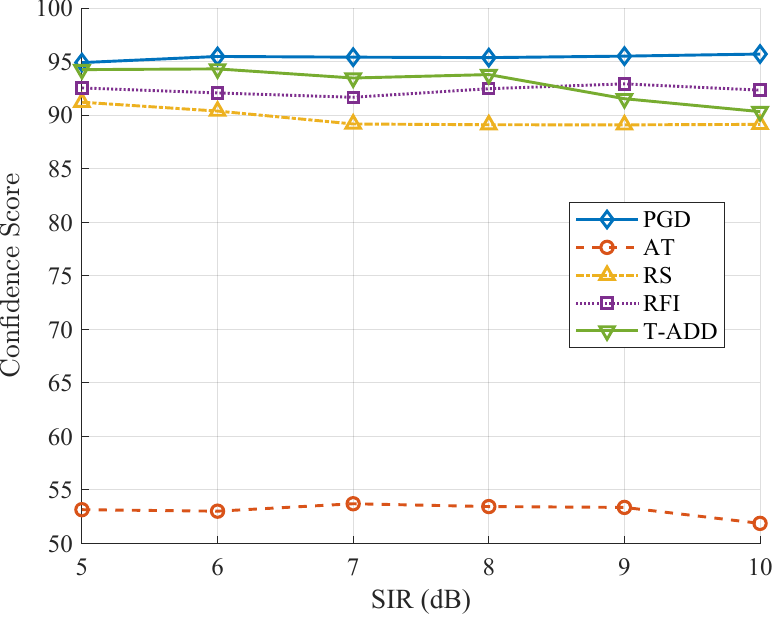}}
\subfigure[]{\label{fig:mim_sig1_xsir_conf}
\includegraphics[width=0.23\linewidth]{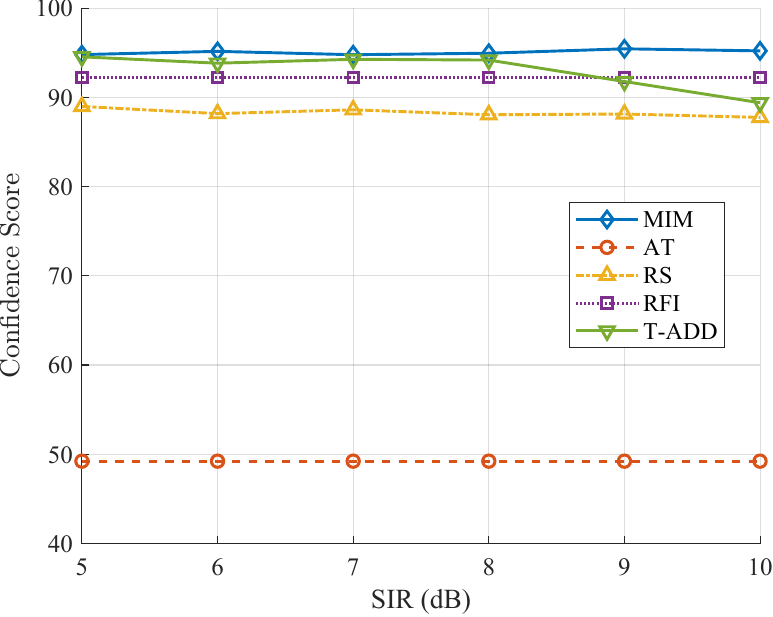}}
\subfigure[]{\label{fig:pgd_sig2_xsir_conf}
\includegraphics[width=0.23\linewidth]{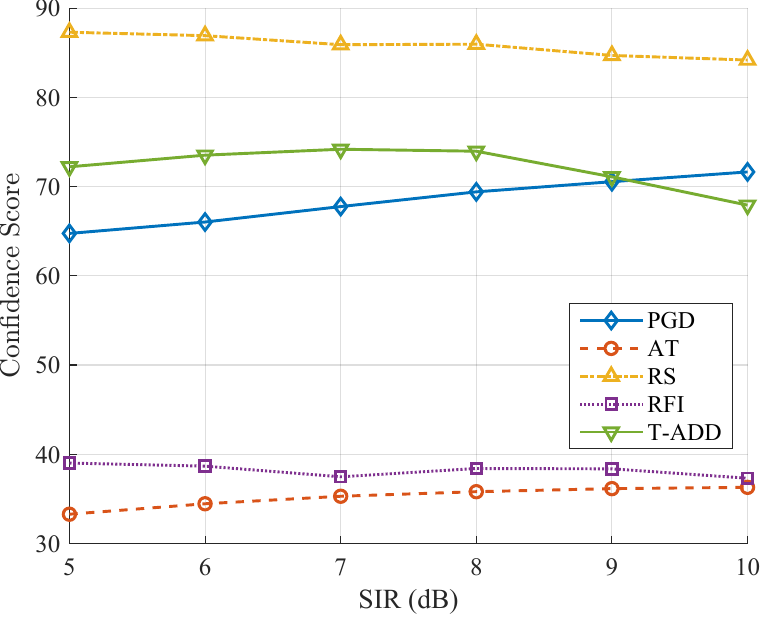}}
\subfigure[]{\label{fig:mim_sig2_xsir_conf}
\includegraphics[width=0.23\linewidth]{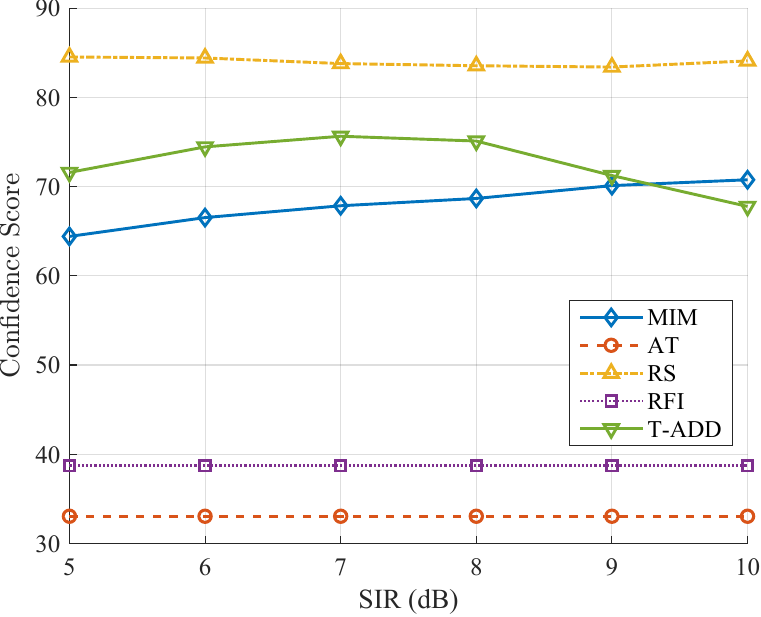}}
\caption{Comparison of different defense methods under PGD and MIM attacks across varying SIRs. Comparison of different defense methods under PGD and MIM attacks across varying SIRs. (a) RMSE under PGD attack with  $L=1$, (b) RMSE under MIM with $L=1$, (c) RMSE under PGD with $L=2$, (d) RMSE under MIM with $L=2$, (e) Accuracy under PGD with $L=1$, (f) Accuracy under MIM with $L=1$, (g) Accuracy under PGD with $L=2$, (h) Accuracy under MIM with $L=2$, (i) Confidence score under PGD with $L=1$, (j) Confidence score under MIM with $L=1$, (k) Confidence score under PGD with $L=2$ and (l) Confidence score under MIM with $L=2$.}

\label{fig-sir}
\end{figure*}
In this experiment, we fix the SNR at \(-1\) dB and set \( K = 1024 \), examining the performance under different SIRs. The angular range is consistent with that described in Section~\ref{section-exp-snr}. For each scenario, 10 clean samples are generated per DOA (or DOA pair), forming the test set. Fig. \ref{fig-sir} presents the results of different methods in varying SIRs. The simulation results indicate that the proposed method maintains strong robustness even under severe interference, consistently delivering \textcolor{black}{consistent} performance compared to other defense methods. 

Specifically, T-ADD demonstrates reliable robustness across different SIRs, reflected in both lower RMSE and higher accuracy. Under low SIR, T-ADD still exhibits favorable robustness and generalization. For instance, in the dual-source scenario under MIM attacks, T-ADD reduces the RMSE by 1.80, corresponding to a 25\% decreasement. In contrast, the AT, RS, and RFI methods exhibit consistently low accuracy across all SIRs. Particularly, AT and RS show minimal improvements, with accuracy remaining below 5.38\% regardless of the SIR. By comparison, T-ADD's accuracy improves significantly as SIR increases, exceeding 66\% at the SIR of 10 dB. This improvement is especially pronounced in the dual-source case, suggesting that the proposed method effectively mitigates misestimation caused by strong interference.

Differences in confidence scores across methods are less pronounced than those observed in RMSE and accuracy. For AT ($L = 1, 2$) and RFI ($L = 2$) exhibit consistently low confidence, reflecting their limited adaptability to complex interference scenarios. RS appears relatively unaffected by varying SIRs; however, when considered alongside its RMSE and accuracy performance, this likely reflects its insufficient robustness, occasionally yielding high confidence scores at incorrect predictions.
\begin{table*}[htbp]
\centering
\footnotesize
\setlength{\tabcolsep}{5pt}
\renewcommand{\arraystretch}{1.16}

\caption{Comparison of RMSE, Accuracy and Confidence Scores of Different Methods under Various Modulation}
\begin{tabular}{|c|c|c|c|c|c|c|c|c|c|c|c|c|c|}
\hline\label{tab:exp-mod}
\multirow{2}{*}{Metric} & \multirow{2}{*}{$L$} & \multirow{2}{*}{Modulation} & \multirow{2}{*}{Attack} 
& \multicolumn{5}{c|}{\textbf{On Adversarial Samples}} 
& \multicolumn{5}{c|}{\textbf{On Clean Samples}} \\
\cline{5-14}
 & & & & No Def & AT & RS & RFI & T-ADD & AT-N & RS-N & RFI-N & T-ADD-N & Baseline \\
\hline
\multirow{16}{*}{RMSE} 
& \multirow{8}{*}{1} 
& \multirow{2}{*}{16QAM} 
& PGD & 5.95 & 1.86 & 2.16 & 13.89 & \textbf{0.87} & \multirow{2}{*}{\textbf{0.00}} & \multirow{2}{*}{\textbf{0.00}} & \multirow{2}{*}{17.85} & \multirow{2}{*}{0.04} & \multirow{2}{*}{\textbf{0.00}} \\
\cline{4-9}
& & & MIM & 6.27 & 4.10 & 9.86 & 19.92 & \textbf{0.69} & & & & & \\
\cline{3-14}
& & \multirow{2}{*}{64QAM} 
& PGD & 6.14 & 1.70 & 2.21 & 13.20 & \textbf{0.87} & \multirow{2}{*}{0.04} & \multirow{2}{*}{\textbf{0.00}} & \multirow{2}{*}{17.72} & \multirow{2}{*}{0.06} & \multirow{2}{*}{0.04} \\
\cline{4-9}
& & & MIM & 6.09 & 4.04 & 10.06 & 19.32 & \textbf{0.66} & & & & & \\
\cline{3-14}
& & \multirow{2}{*}{256QAM} 
& PGD & 6.30 & 1.67 & 2.18 & 12.48 & \textbf{0.90} & \multirow{2}{*}{0.04} & \multirow{2}{*}{\textbf{0.00}} & \multirow{2}{*}{18.23} & \multirow{2}{*}{0.06} & \multirow{2}{*}{\textbf{0.00}} \\
\cline{4-9}
& & & MIM & 6.22 & 4.00 & 8.44 & 20.04 & \textbf{0.69} & & & & & \\
\cline{3-14}
& & \multirow{2}{*}{\textcolor{black}{OFDM}}
& \textcolor{black}{PGD} & \textcolor{black}{5.01} & \textcolor{black}{3.48} & \textcolor{black}{6.28} & \textcolor{black}{17.60} & \textcolor{black}{\textbf{0.88}} & \multirow{2}{*}{\textcolor{black}{0.04}} & \multirow{2}{*}{\textcolor{black}{\textbf{0.00}}} & \multirow{2}{*}{\textcolor{black}{17.71}} & \multirow{2}{*}{\textcolor{black}{0.06}} & \multirow{2}{*}{\textcolor{black}{\textbf{0.00}}} \\
\cline{4-9}
& & & \textcolor{black}{MIM} & \textcolor{black}{4.18} & \textcolor{black}{3.37} & \textcolor{black}{6.33} & \textcolor{black}{14.92} & \textcolor{black}{\textbf{0.70}} & & & & &  \\

\cline{2-14}
& \multirow{8}{*}{2} 
& \multirow{2}{*}{16QAM} 
& PGD & 3.49 & 4.59 & 6.03 & 4.19 & \textbf{0.93} & \multirow{2}{*}{2.36} & \multirow{2}{*}{4.17} & \multirow{2}{*}{2.94} & \multirow{2}{*}{0.22} & \multirow{2}{*}{\textbf{0.20}} \\
\cline{4-9}
& & & MIM & 3.60 & 4.73 & 6.02 & 4.26 & \textbf{0.93} & & & & & \\
\cline{3-14}
& & \multirow{2}{*}{64QAM} 
& PGD & 3.39 & 4.73 & 5.74 & 4.23 & \textbf{0.99} & \multirow{2}{*}{2.34} & \multirow{2}{*}{4.19} & \multirow{2}{*}{2.95} & \multirow{2}{*}{0.24} & \multirow{2}{*}{\textbf{0.21}} \\
\cline{4-9}
& & & MIM & 3.49 & 4.58 & 5.71 & 4.20 & \textbf{0.95} & & & & & \\
\cline{3-14}
& & \multirow{2}{*}{256QAM} 
& PGD & 3.41 & 4.65 & 5.90 & 4.13 & \textbf{0.95} & \multirow{2}{*}{2.29} & \multirow{2}{*}{4.14} & \multirow{2}{*}{2.95} & \multirow{2}{*}{0.22} & \multirow{2}{*}{\textbf{0.19}} \\
\cline{4-9}
& & & MIM & 3.48 & 4.70 & 5.95 & 4.20 & \textbf{0.90} & & & & & \\
\cline{3-14}
& & \multirow{2}{*}{\textcolor{black}{OFDM}} 
& \textcolor{black}{PGD} & \textcolor{black}{3.32} & \textcolor{black}{4.35} & \textcolor{black}{5.13} & \textcolor{black}{4.16} & \textcolor{black}{\textbf{1.36}} & \multirow{2}{*}{\textcolor{black}{2.21}} & \multirow{2}{*}{\textcolor{black}{4.01}} & \multirow{2}{*}{\textcolor{black}{2.99}} & \multirow{2}{*}{\textcolor{black}{0.05}} & \multirow{2}{*}{\textcolor{black}{\textbf{0.04}}}\\
\cline{4-9}
& & & \textcolor{black}{MIM} & \textcolor{black}{3.41} & \textcolor{black}{4.27} & \textcolor{black}{5.15} & \textcolor{black}{4.38} & \textcolor{black}{\textbf{1.19}} & & & & & \\

\hline
\multirow{16}{*}{Accuracy} 
& \multirow{8}{*}{1} 
& \multirow{2}{*}{16QAM} 
& PGD & 3.30 & 83.52 & 72.68 & 68.60 & \textbf{97.96} & \multirow{2}{*}{\textbf{100.00}} & \multirow{2}{*}{\textbf{100.00}} & \multirow{2}{*}{94.51} & \multirow{2}{*}{\textbf{100.00}} & \multirow{2}{*}{\textbf{100.00}} \\
\cline{4-9}
& & & MIM & 1.73 & 1.26 & 0.63 & 4.40 & \textbf{99.22} & & & & & \\
\cline{3-14}
& & \multirow{2}{*}{64QAM} 
& PGD & 3.14 & 83.52 & 72.06 & 68.29 & \textbf{97.33} & \multirow{2}{*}{\textbf{100.00}} & \multirow{2}{*}{\textbf{100.00}} & \multirow{2}{*}{94.66} & \multirow{2}{*}{\textbf{100.00}} & \multirow{2}{*}{\textbf{100.00}} \\
\cline{4-9}
& & & MIM & 1.26 & 1.26 & 0.94 & 4.40 & \textbf{98.74} & & & & & \\
\cline{3-14}
& & \multirow{2}{*}{256QAM} 
& PGD & 3.45 & 84.14 & 72.21 & 68.76 & \textbf{97.02} & \multirow{2}{*}{\textbf{100.00}} & \multirow{2}{*}{\textbf{100.00}} & \multirow{2}{*}{93.72} & \multirow{2}{*}{\textbf{100.00}} & \multirow{2}{*}{\textbf{100.00}} \\
\cline{4-9}
& & & MIM & 2.51 & 2.83 & 0.00 & 3.61 & \textbf{98.90} & & & & & \\
\cline{3-14}
& & \multirow{2}{*}{\textcolor{black}{OFDM}} 
& \textcolor{black}{PGD} & \textcolor{black}{5.81} & \textcolor{black}{0.94} & \textcolor{black}{7.85} & \textcolor{black}{10.52} & \textcolor{black}{\textbf{92.27}} & \multirow{2}{*}{\textcolor{black}{\textbf{100.00}}} & \multirow{2}{*}{\textcolor{black}{\textbf{100.00}}} & \multirow{2}{*}{\textcolor{black}{94.51}} & \multirow{2}{*}{\textcolor{black}{\textbf{100.00}}} & \multirow{2}{*}{\textcolor{black}{\textbf{100.00}}} \\
\cline{4-9}
& & & \textcolor{black}{MIM} & \textcolor{black}{3.92} & \textcolor{black}{2.51} & \textcolor{black}{2.67} & \textcolor{black}{5.97} & \textcolor{black}{\textbf{99.22}} & & & & & \\

\cline{2-14}
& \multirow{8}{*}{2} 
& \multirow{2}{*}{16QAM} 
& PGD & 0.00 & 0.36 & 0.00 & 2.73 & \textbf{97.64} & \multirow{2}{*}{56.91} & \multirow{2}{*}{0.00} & \multirow{2}{*}{15.09} & \multirow{2}{*}{\textbf{100.00}} & \multirow{2}{*}{\textbf{100.00}} \\
\cline{4-9}
& & & MIM & 0.00 & 0.73 & 0.00 & 2.00 & \textbf{97.64} & & & & & \\
\cline{3-14}
& & \multirow{2}{*}{64QAM} 
& PGD & 0.00 & 0.73 & 0.00 & 1.64 & \textbf{96.91} & \multirow{2}{*}{57.09} & \multirow{2}{*}{0.00} & \multirow{2}{*}{12.36} & \multirow{2}{*}{\textbf{100.00}} & \multirow{2}{*}{\textbf{100.00}} \\
\cline{4-9}
& & & MIM & 0.00 & 0.55 & 0.00 & 2.18 & \textbf{97.45} & & & & & \\
\cline{3-14}
& & \multirow{2}{*}{256QAM} 
& PGD & 0.00 & 0.91 & 0.00 & 2.18 & \textbf{98.36} & \multirow{2}{*}{59.27} & \multirow{2}{*}{0.00} & \multirow{2}{*}{13.82} & \multirow{2}{*}{\textbf{100.00}} & \multirow{2}{*}{\textbf{100.00}} \\
\cline{4-9}
& & & MIM & 0.00 & 1.27 & 0.00 & 2.18 & \textbf{97.27} & & & & & \\
\cline{3-14}
& & \multirow{2}{*}{\textcolor{black}{OFDM}} 
& \textcolor{black}{PGD} & \textcolor{black}{0.00} & \textcolor{black}{1.58} & \textcolor{black}{0.00} & \textcolor{black}{1.70} & \textcolor{black}{\textbf{88.97}} & \multirow{2}{*}{\textcolor{black}{62.06}} & \multirow{2}{*}{\textcolor{black}{0.00}} & \multirow{2}{*}{\textcolor{black}{12.48}} & \multirow{2}{*}{\textcolor{black}{\textbf{100.00}}} & \multirow{2}{*}{\textcolor{black}{\textbf{100.00}}} \\
\cline{4-9}
& & & \textcolor{black}{MIM} & \textcolor{black}{0.00} & \textcolor{black}{1.09} & \textcolor{black}{0.00} & \textcolor{black}{3.63} & \textcolor{black}{\textbf{92.97}} & & & & & \\

\hline
\multirow{16}{*}{Confidence} 
& \multirow{8}{*}{1} 
& \multirow{2}{*}{16QAM} 
& PGD & \textbf{98.46} & 74.88 & 95.68 & 95.25 & 98.36 & \multirow{2}{*}{75.83} & \multirow{2}{*}{99.88} & \multirow{2}{*}{93.97} & \multirow{2}{*}{\textbf{100.00}} & \multirow{2}{*}{\textbf{100.00}} \\
\cline{4-9}
& & & MIM & 97.88 & 59.71 & 94.30 & 93.51 & \textbf{99.03} & & & & & \\
\cline{3-14}
& & \multirow{2}{*}{64QAM} 
& PGD & 98.49 & 74.89 & 96.23 & 94.46 & \textbf{98.54} & \multirow{2}{*}{75.82} & \multirow{2}{*}{99.94} & \multirow{2}{*}{94.04} & \multirow{2}{*}{99.99} & \multirow{2}{*}{\textbf{100.00}} \\
\cline{4-9}
& & & MIM & 98.23 & 60.11 & 93.61 & 93.80 & \textbf{98.89} & & & & & \\
\cline{3-14}
& & \multirow{2}{*}{256QAM} 
& PGD & 98.42 & 75.51 & 96.29 & 95.37 & \textbf{98.48} & \multirow{2}{*}{75.94} & \multirow{2}{*}{49.99} & \multirow{2}{*}{94.04} & \multirow{2}{*}{99.91} & \multirow{2}{*}{\textbf{100.00}} \\
\cline{4-9}
& & & MIM & 98.32 & 59.82 & 93.93 & 93.67 & \textbf{98.99} & & & & & \\
\cline{3-14}
& & \multirow{2}{*}{\textcolor{black}{OFDM}} 
& \textcolor{black}{PGD} & \textcolor{black}{98.28} & \textcolor{black}{72.95} & \textcolor{black}{95.97} & \textcolor{black}{95.37} & \textcolor{black}{\textbf{98.48}} & \multirow{2}{*}{\textcolor{black}{76.12}} & \multirow{2}{*}{\textcolor{black}{100.00}} & \multirow{2}{*}{\textcolor{black}{94.42}} & \multirow{2}{*}{\textcolor{black}{99.91}} & \multirow{2}{*}{\textcolor{black}{\textbf{99.98}}} \\
\cline{4-9}
& & & \textcolor{black}{MIM} & \textcolor{black}{98.32} & \textcolor{black}{59.82} & \textcolor{black}{93.93} & \textcolor{black}{93.67} & \textcolor{black}{\textbf{98.99}} & & & & & \\

\cline{2-14}
& \multirow{8}{*}{2} 
& \multirow{2}{*}{16QAM} 
& PGD & 73.44 & 34.52 & 86.31 & 34.16 & \textbf{90.58} & \multirow{2}{*}{24.03} & \multirow{2}{*}{73.94} & \multirow{2}{*}{28.98} & \multirow{2}{*}{94.68} & \multirow{2}{*}{\textbf{94.75}} \\
\cline{4-9}
& & & MIM & 76.19 & 30.89 & 86.05 & 30.58 & \textbf{90.73} & & & & & \\
\cline{3-14}
& & \multirow{2}{*}{64QAM} 
& PGD & 73.44 & 34.92 & 87.72 & 30.76 & \textbf{90.41} & \multirow{2}{*}{24.06} & \multirow{2}{*}{73.27} & \multirow{2}{*}{29.82} & \multirow{2}{*}{94.34} & \multirow{2}{*}{\textbf{94.47}} \\
\cline{4-9}
& & & MIM & 73.77 & 31.06 & 87.51 & 32.46 & \textbf{91.02} & & & & & \\
\cline{3-14}
& & \multirow{2}{*}{256QAM} 
& PGD & 74.97 & 34.23 & 85.47 & 32.71 & \textbf{91.15} & \multirow{2}{*}{24.06} & \multirow{2}{*}{72.52} & \multirow{2}{*}{29.17} & \multirow{2}{*}{94.50} & \multirow{2}{*}{\textbf{94.51}} \\
\cline{4-9}
& & & MIM & 75.15 & 30.95 & 87.25 & 32.77 & \textbf{91.18} & & & & & \\
\cline{3-14}
& & \multirow{2}{*}{\textcolor{black}{OFDM}} 
& \textcolor{black}{PGD} & \textcolor{black}{74.97} & \textcolor{black}{34.23} & \textcolor{black}{85.47} & \textcolor{black}{32.71} & \textcolor{black}{\textbf{91.15}} & \multirow{2}{*}{\textcolor{black}{24.10}} & \multirow{2}{*}{\textcolor{black}{83.06}} & \multirow{2}{*}{\textcolor{black}{29.35}} & \multirow{2}{*}{\textcolor{black}{96.34}} & \multirow{2}{*}{\textcolor{black}{\textbf{96.40}}} \\
\cline{4-9}
& & & \textcolor{black}{MIM} & \textcolor{black}{75.15} & \textcolor{black}{30.95} & \textcolor{black}{87.25} & \textcolor{black}{32.77} & \textcolor{black}{\textbf{91.18}} & & & & & \\

\hline
\end{tabular}
\end{table*}

\subsubsection{Comparison under Different Modulations}

To evaluate the generalization capability of the proposed method, we conduct testing on modulation types not seen during training, including 16QAM, 64QAM, \textcolor{black}{OFDM} and 256QAM. In this experiment, the SNR is fixed at 6 dB. For the test data, the single-source scenario spans a DOA range of $[-45^\circ, 45^\circ]$, with 7 clean samples generated per angle. In the dual-source scenario, the DOA range is set to $[-30^\circ, 30^\circ]$, with an angular separation of $6^\circ$, and 10 clean samples generated per DOA pair. Table~\ref{tab:exp-mod} summarizes the performance of different methods under these modulation scheme\textcolor{black}{s}. ``On Adversarial Samples'' and ``On Clean Samples'' refer to performance on adversarial and clean samples, respectively, while ``No Def'' and ``Baseline'' represent the performance of the baseline DOA model on adversarial and clean samples.

Experimental results demonstrate that the proposed T-ADD method consistently exhibits strong robustness across all \textcolor{black}{four} modulation schemes and significantly outperforms other defense methods. Notably, T-ADD achieves the highest accuracy and the lowest RMSE on adversarial samples, while maintaining performance on clean samples close to that of the Baseline.

For example, \textcolor{black}{when $L=1$}, 256QAM under MIM attack, T-ADD reduces the RMSE by 5.53 compared to No Def, indicating superior resistance to adversarial perturbations. In the dual-source case, 64QAM scenario, T-ADD achieves RMSEs of 3.95 and 2.10 lower than RS and AT, respectively, with only a 0.03 increment from the Baseline, highlighting excellent generalization ability. In contrast, while other methods may perform reasonably well under single-source PGD attacks, their performance degrades significantly under MIM attacks or in dual-source settings. In the case of $L=1$, 16QAM, and MIM attacks, T-ADD achieves an accuracy of 99.22\%, identical to its accuracy on clean samples. \textcolor{black}{Under OFDM, our method also achieves significantly superior defense performance compared to other defense methods.}

In terms of confidence scores, both T-ADD and T-ADD-N maintain confidence scores close to the Baseline across different modulation schemes. Other methods, however, generally suffer significant drops in confidence score. Notably, in the $L=2$ setting, the confidence scores of AT-N and RFI-N decrease by more than 64\% compared to the Baseline, clearly exposing their poor generalization to unseen modulation schemes.

\begin{table*}[tbp]
\centering
\footnotesize
\setlength{\tabcolsep}{6pt}  
\caption{{Comparison of Different Methods under Sparse Linear Array}}
\begin{tabular}{|c|c|c|c|c|c|c|c|c|c|c|c|c|}
\hline\label{tab:sla-exp}
\multirow{2}{*}{\(L\)} & \multirow{2}{*}{Metric} & \multirow{2}{*}{Attack} 
& \multicolumn{5}{c|}{\textbf{On Adversarial Samples}} 
& \multicolumn{5}{c|}{\textbf{On Clean Samples}} \\
\cline{4-13}
 &  &  & No Def & AT & RS & RFI & T-ADD & AT-N & RS-N & RFI-N & T-ADD-N & Baseline \\
\hline
\multirow{6}{*}{1} 
& \multirow{2}{*}{RMSE} 
& PGD & 8.48 & 13.19 & 10.14 & 14.98 & \textbf{1.95} & \multirow{2}{*}{0.52} & \multirow{2}{*}{\textbf{0.51}} & \multirow{2}{*}{0.53} & \multirow{2}{*}{0.58} & \multirow{2}{*}{0.51} \\
\cline{3-8}
& & MIM & 9.83 & 12.10 & 11.40 & 13.76 & \textbf{1.90} & & & & & \\
\cline{2-13}
& \multirow{2}{*}{Acc} 
& PGD & 0.17 & 0.00 & 0.00 & 0.00 & \textbf{90.67} & \multirow{2}{*}{\textbf{100.00}} & \multirow{2}{*}{82.05} & \multirow{2}{*}{28.11} & \multirow{2}{*}{99.83} & \multirow{2}{*}{100.00} \\
\cline{3-8}
& & MIM & 1.50 & 0.00 & 0.00 & 0.00 & \textbf{88.67} & & & & & \\
\cline{2-13}
& \multirow{2}{*}{Conf} 
& PGD & \textbf{97.17} & 27.93 & 89.78 & 28.11 & 91.83 & \multirow{2}{*}{24.43} & \multirow{2}{*}{\textbf{100.00}} & \multirow{2}{*}{27.46} & \multirow{2}{*}{90.98} & \multirow{2}{*}{88.93} \\
\cline{3-8}
& & MIM & \textbf{96.85} & 26.31 & 88.79 & 27.58 & 90.76 & & & & & \\
\hline
\multirow{6}{*}{2} 
& \multirow{2}{*}{RMSE} 
& PGD & 4.91 & 23.16 & 36.50 & 24.40 & \textbf{0.91} & \multirow{2}{*}{15.30} & \multirow{2}{*}{19.23} & \multirow{2}{*}{21.66} & \multirow{2}{*}{\textbf{1.00}} & \multirow{2}{*}{0.52} \\
\cline{3-8}
& & MIM & 5.29 & 23.28 & 34.63 & 24.65 & \textbf{0.81} & & & & & \\
\cline{2-13}
& \multirow{2}{*}{Acc} 
& PGD & 80.00 & 3.42 & 0.00 & 0.00 & \textbf{92.07} & \multirow{2}{*}{3.24} & \multirow{2}{*}{5.23} & \multirow{2}{*}{0.00} & \multirow{2}{*}{\textbf{97.14}} & \multirow{2}{*}{99.46} \\
\cline{3-8}
& & MIM & 73.15 & 2.70 & 0.18 & 0.00 & \textbf{96.40} & & & & & \\
\cline{2-13}
& \multirow{2}{*}{Conf} 
& PGD & 82.37 & 26.03 & 81.22 & 23.80 & \textbf{98.42} & \multirow{2}{*}{14.16} & \multirow{2}{*}{32.48} & \multirow{2}{*}{64.05} & \multirow{2}{*}{51.07} & \multirow{2}{*}{\textbf{97.44}} \\
\cline{3-8}
& & MIM & 84.17 & 31.63 & 81.55 & 23.98 & \textbf{97.83} & & & & & \\
\hline
\end{tabular}
\end{table*}
\begin{table}[tbp]
\caption{Comparison of Inference Latency of Different Methods}
\centering
\label{tab:latency}
\footnotesize
\setlength{\tabcolsep}{2.25pt}  
\begin{tabular}{|c|c|c|c|c|c|c|}
\hline
MUSIC & ESPRIT & T-ADD & AT & RS & RFI & CNN \\
\hline
 20.75 ms& 0.62 ms & 14.06 ms & 0.40 ms& 42.39 ms& 0.40 ms& 0.40 ms\\
\hline
\end{tabular}
\end{table}

\subsubsection{Comparison under Sparse Linear Array}
To evaluate the robustness of the proposed method under non-uniform array configurations, we further conduct simulations on a SLA. Specifically, an SLA comprising 8 elements is constructed, where the array positions are defined as $\bm{\Delta} = [0, 1, 2, 5, 6, 9, 10, 11]$. The same baseline model architecture used in previous simulations is retained for training, although it was originally designed for ULAs and not specifically tailored for SLAs. Therefore, a certain performance degradation might be expected in sparse configurations.

During training, the angular range is set to $[-45^\circ, 45^\circ]$ for single-source cases and $[-30^\circ, 30^\circ]$ for dual-source scenario. The number of generated samples per DOA is consistent with Section~\ref{sec:dataset-setup}. The SNR of this evaluation is fixed at $-2\ \text{dB}$. For the single-source case, the DOA range is set to $[-29.5^\circ, 29.5^\circ]$ with a resolution of $1^\circ$, and 10 samples are generated for each angle. In the dual-source scenario, the range is $[-19.5^\circ, 19.5^\circ]$ with an angular separation of $3^\circ$, generating 15 samples per DOA pair.

The experimental results, presented in Table.~\ref{tab:sla-exp}, show the RMSE, accuracy, and confidence scores of various methods under the SLA configuration. The results indicate that the proposed method retains strong generalization ability and significantly enhances model robustness in sparse array scenarios compared to other defense methods.

Specifically, T-ADD consistently demonstrates superior performance in terms of both RMSE and accuracy. Under the $L=2$ MIM attack setting, T-ADD achieves an RMSE of 0.81 and an accuracy of 96.40\%, clearly outperforming the other methods. Moreover, the RMSE and accuracy differences between T-ADD and the Baseline are only 0.48 and 2.32\%, suggesting that the proposed method maintains the performance which is close to Baseline  even under SLA.

Although RS-N and AT-N outperform T-ADD-N in certain metrics under $L=1$, their generalization performance degrades significantly when $L=2$. For instance, AT-N’s RMSE increases by 14.78 compared to the Baseline, and RS-N’s accuracy drops by 94.23\%. In terms of confidence scores, AT and RFI produce significantly lower confidence than the Baseline, whereas only RS and T-ADD maintain high confidence scores—further validating the robustness of our approach.


\subsection{Computational Complexity}

We provide the computational complexity of the Def-Transformer model. The number of floating point operations (FLOPs) by the Def-Transformer required is $62.56\text{M}$, and the total number of parameters is $0.88\text{M}$. In comparison, the baseline CNN model used for DOA estimation requires $20.8\text{M}$ FLOPs, but has a substantially larger parameter count of $33.4\text{M}$. All computations are performed on a platform equipped with an Intel Xeon Gold 6430 CPU and an NVIDIA GeForce RTX 4090 GPU. FLOPs and parameter counts are measured utilizing the Python library \texttt{thop}.

Furthermore, we evaluate and compare the average inference time of various methods, including the baseline CNN, T-ADD, AT, RS, RFI, and the classic subspace-based algorithms MUSIC and ESPRIT. For MUSIC, the angular search step is set to $0.5^{\circ}$. The inference latency for each method is obtained by averaging over 2000 independent runs. The results are summarized in Table~\ref{tab:latency}.
As shown in the table, the RS method exhibits the highest inference latency. The inference time of T-ADD lies between that of MUSIC and ESPRIT. For the AT and RFI methods, the inference latency is identical to that of the baseline model, as their additional computational overhead primarily stems from the adversarial sample generation during training, rather than from inference. Although the FLOPs of T-ADD are roughly three times those of the baseline CNN, its inference latency increases more significantly. This discrepancy may be attributed to the fragmented operations and non-contiguous memory access patterns in the transformer architecture, which reduce hardware utilization efficiency during execution. \textcolor{black}{On a computer with Intel I7-9700 CPU, 24 GB RAM, and NVIDIA RTX 2070 SUPER GPU, the training time of T-ADD is: 1.18 hours for the single-source scenario and 9.76 hours for the dual-source scenario.}

\section{Conclusion}
In this paper, we proposed T-ADD, a defense framework designed to enhance adversarial robustness in DOA estimation based on deep learning. Our approach leverages a transformer model to capture the intricate mapping between adversarial and clean samples, thereby effectively suppressing adversarial perturbations in input data.
Extensive experiments under two representative adversarial attacks demonstrate that, compared with three existing state-of-the-art defense methods, the proposed approach achieves a superior trade-off between robustness and  performance on clean sample, while also exhibiting improved robustness \textcolor{black}{under different} modulation schemes, attacks and array structures.

\textcolor{black}{While T-ADD shows promising results under far-field and narrow band assumptions, its performance in near-field scenarios and with other array structures such as nested arrays remains to be fully explored.} Future work will focus on extending the proposed framework to more challenging scenarios. \textcolor{black}{For example, the unseen type of attacks like} black-box attacks and transfer-based attacks will be explored in practical deployment environments. We also aim to further explore its generalization capabilities in multi-source signal scenarios\textcolor{black}{, near field or wide-band signal, 2D-DOA estimation and more advanced antenna arrays.}

\bibliographystyle{IEEEtran}
\bibliography{citationlist}

\end{document}